\documentclass[usenatbib, useAMS]{mn2e}
\usepackage{graphicx, subfigure, amsmath }

\title[Galaxy gas flows and metal budget] {Galaxy gas flows inferred from a detailed, spatially resolved
metal budget}
\author[F. Belfiore et al.] {F.~Belfiore,$^{1,2}$ R.~Maiolino$^{1,2}$ and M.~Bothwell$^{1,2}$  
\\$^1$ Kavli Institute for Cosmology, University of Cambridge, Madingley Road, CB3 0HA Cambridge, UK 
\\$^2$ Cavendish Laboratory, University of Cambridge, 19 J. J. Thomson Avenue, Cambridge, CB3 0HE, UK}

\begin{document}

\maketitle
  \begin{abstract}
We use the most extensive integral field spectroscopic map of a local galaxy, NGC 628, combined with gas
and stellar mass surface density maps, to study the distribution of metals in this galaxy out to 3 effective radii ($\rm R_e$). At each galactocentric distance, we compute the metal budget and thus constrain the mass of metals lost. 
We find that in the disc about 50\% of the metals have been lost throughout the lifetime of the galaxy. The fraction of metals lost is higher in the bulge ($\sim$70\%) and decreases towards the outer disc ($\rm \sim 3 \ R_e$). In contrast to studies based on the gas kinematics, which are only sensitive to ongoing outflow events, our metal budget analysis enables us to infer the average outflow rate during the galaxy lifetime. By using simple physically motivated models of chemical evolution we can fit the observed metal budget at most radii with an average outflow loading factor of order unity, thus clearly demonstrating the importance of outflows in the evolution of disc galaxies 
of this mass range ($\rm log(M_\star/M_\odot) \sim 10)$.
The observed gas phase metallicity is higher than expected from the metal budget and suggests late-time accretion of enriched gas, likely raining onto the disc from the metal-enriched halo.

\end{abstract}

\section{Introduction}

Current models of galaxy evolution predict that star formation and consequent metal enrichment in galaxies are regulated by a combination of gas inflows and outflows. In this framework, the enrichment of the intergalactic medium (IGM) is mediated by metal-rich outflows produced by galactic winds, especially during the early, more active phases of galaxy evolution \citep{Oppenheimer2006,  Scannapieco2006, Shen2010, Oppenheimer2012}. 
Evidence for ongoing metal-enriched outflows has been obtained both through absorption line studies of quasars whose lines
of sight cross the vicinity of nearby galaxies \citep{Tumlinson2011, Tumlinson2013, Werk2013, Bregman2013} and through
direct imaging of outflowing ionised or neutral/molecular gas at low and high redshift \citep{Heckmann2000,
Martin2005,Veilleux2005, Martin2006, Tremonti2007, Feruglio2010, Sturm2011, Cano-Diaz2012, Maiolino2012, Cicone2014, Cazzoli2014,
Arribas2014, Cicone2015}. However, these observations are only probing ongoing events. A global understanding of the effect of outflows on the cosmic star formation history and IGM enrichment is only possible if we can gain insight into gas flows over the entire evolutionary history of galaxies.
The content and distribution of metals in galaxies can provide such information.

Metals are produced by stellar nucleosynthesis and act as a fossil record of a galaxy's star formation history and gas flows. Simple analytical models relating metal abundances to star formation rate (SFR) and gas flow rates have been presented by several authors \citep{Erb2008, Lilly2013, Dayal2013, Peng2014, Ascasibar2014} with the aim of explaining the scaling relations observed in statistical samples of galaxies. More detailed chemical evolutionary models have been successfully exploited to reproduce the observed chemical abundances, gas content and SFR of the solar neighbourhood, the Milky Way or nearby disc galaxies \citep{Colavitti2009, Marcon-Uchida2010, Spitoni2011}. 

The common aim of chemical evolutionary models is to use chemical abundances observed today to infer the gas flow history
and star formation history of the system. For example, in order to reproduce the chemical abundances of the Milky Way, a two-stage inflow model is
generally invoked. The bulge is assumed to have formed at early times during a first inflow phase, while the disc is
assumed to form inside-out during a second inflow phase \citep{Matteucci1989, Boissier1999, Naab2006, Williams2009}. Hydrodynamical
simulations of disc galaxies generally agree with the inside-out disc formation model \citep{Pilkington2012a}. Chemical
abundances can also provide clues on the relative importance of additional evolutionary processes in galaxies, including stellar migration, gas flows within the disc and large scale galactic fountains.

Current large integral field spectroscopy (IFS) galaxy surveys (eg. CALIFA, \citealt{Sanchez2012a}; SAMI, \citealt{Croom2012}; MaNGA, \citealt{Bundy2015}) offer great potential for extending the study of resolved chemical abundances to much larger galaxy samples. Using optical IFS data, the chemical abundance of the interstellar medium (ISM) can be derived by analysing gas emission lines, while insight into the metallicity of the stellar population can be obtained using Lick indices, or applying spectral decomposition techniques \citep{Sanchez-Blazquez2014, CidFernandes2011}. The CALIFA survey has recently delivered important insights on the distribution of metals within star-forming galaxies, providing the strongest evidence so far for a universal gas phase abundance gradient \citep{Sanchez2014} out to 2 effective radii ($\rm R_{e}$), as already suggested in previous work \citep{Vila-Costas1992, Bresolin2009, Bresolin2012}.

However, observations of resolved gas metallicity alone cannot be uniquely related to specific
chemical evolutionary models and, unless measurements of the gas and stellar content are also available,
the information provided by the gas metallicity is highly degenerate between inflows and outflows.

This work presents one of the first attempts to combine chemical abundances from IFS with gas mass measurements from
millimetre/radio observations for a single galaxy (NGC 628) on the same spatial scales, along with stellar mass surface
density and stellar metallicities. We are therefore able to constrain the metal budget both as a function of radius, and for the galaxy as a whole, out to $\rm 3 \ R_e$. Since we have access to information on radial gradients of the relevant physical quantitates, we obtain an overall metal budget free from aperture effects, which affected previous work based on metallicities estimated from observations of galaxy central regions (for example from the SDSS galaxy sample, \citealt{York2000, Abazajian2009}).

Empirical studies of the metal budget in the z = 0 Universe, presented by several authors, agree in concluding that the total mass of metals produced by galaxies cannot be accounted for by the metals observed in stars and the ISM \citep{Pettini1999, Ferrara2005, Bouche2007, Peeples2011, Zahid2012, Peeples2014}. Large amounts of metals (between $35 \%$ and $90 \%$) must be residing in the intergalactic medium, but also in the diffuse halo gas and in the circumgalactic medium (CGM).
Although recent observational efforts have led to a better characterisation of the metallicity of the gas in the hot halo phase \citep{Tumlinson2011, Tumlinson2013}, a robust metal budget for these phases is still missing.

Cosmological hydrodynamical simulations which include chemical evolution in a self-consistent way generally agree with the empirical studies above, and highlight that the fraction of missing metals is strongly dependant of the sub-grid feedback prescription \citep{Wiesma2011, Pilkington2012a}. 

The aim of this work is to investigate in detail how metals are lost from the galaxy, hence the severity of the `missing
metal problem'. This information also provides tight constraints on the outflow loading factor (i.e. the ratio between SFR and outflow rate) {\it averaged} over the galaxy lifetime. The results are expected to provide simulators with new constraints on the net effect of gas flows in disc galaxies on resolved scales.

The paper is structured as follows. In Sec. \ref{over} we give a brief overview of the observational data used, in Sec. \ref{analysis} we summarise how we use the data to derive gas phase metallicity, gas and stellar mass surface densities and their radial gradients. In Sec. \ref{chem_evo} we summarise the chemical evolution framework, while in Sec. \ref{results} we present the results on the metal budget and the modelling of the outflow loading factor in NGC 628. In Sec. \ref{dis} and \ref{concl} we present the discussion and conclusions.

\section{Overview of the data} \label{over}

\begin{figure} 
\includegraphics[width=0.5\textwidth, trim=120 80 80 300, clip]{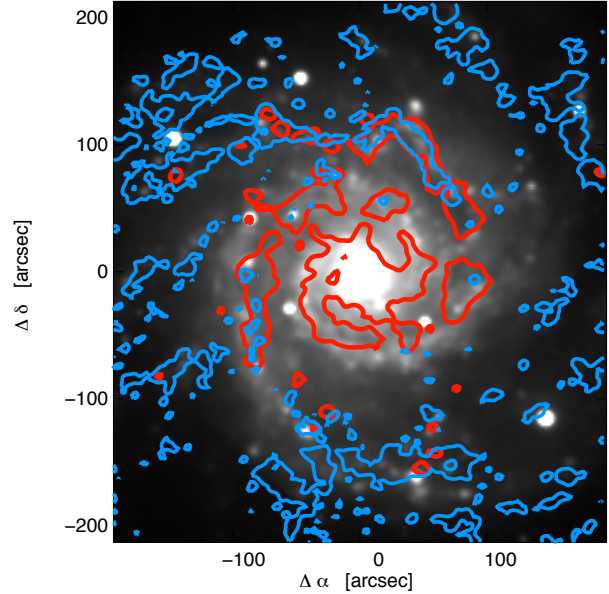}
\caption{Optical image of NGC 628 (r$'$ band from SDSS) with superimposed contours of H\textsc{i} (from the THINGS survey, in blue) and CO surface brightness (from the HERACLES survey, in red). The red contours correspond to a molecular hydrogen surface mass density of $\mathrm{10 \ M_{\odot} pc^{-2}}$, while the blue H\textsc{i} contour correspond to a neutral hydrogen surface mass density of $\mathrm{12 \ M_{\odot} pc^{-2} }$.}
\label{pic1}
\end{figure}

\begin{table}
\caption{General properties of NGC 628}
\label{gal_properties}

\centering
\begin{tabular}{ l c r }

 \hline 
  Property & Value  \\
  \hline 
  Name & NGC 628, M 74, UGC 1149 & \\
  RA & $00^{h}59^{m}50.1^{s}$ \\
  Dec & $-07^{\circ} 34' 41.0''$  \\
  Type& SAc \\
  Distance (adopted) & 7.3 Mpc  \\
  i  & $7^{\circ}$  \\
  $R_{24}$  & $\rm 4.88' = 10.3 \ kpc $\\
  $R_{e} $ & $\rm 67'' = 2.4 \ kpc$ \\
  scale & $\rm 36 \ pc  \ arcsec^{-1}$
  
\end{tabular}
\end{table}

NGC 628 is a nearby ($\rm D \approx 7.3 \ \mathrm{Mpc}, z = 0.00219 $) nearly face-on ($\rm i \approx 7 ^{\circ} $) spiral galaxy (Fig. \ref{pic1}). Table \ref{gal_properties} summarises some of its basic properties. NGC 628 is a good example of an isolated galaxy (no encounter in the last Gyr; see \citealt{Kamphuis1992}), displaying typical grand-design structure.
The following sections describe the observations of  NGC 628 that have been used in this work.

\subsection{Integral field spectroscopy}
In this study we make use of IFS observations of NGC 628 taken as part of the PINGS survey \citep{Rosales-Ortega2010, Sanchez2011, Rosales-Ortega2011}. The survey was carried out using the 3.5-m telescope at the Calar Alto observatory, with the Potsdam Multi-Aperture Spectrograph in PPAK mode \citep{Kelz2006}. The PPAK fibre bundle corresponds to a field of view of $75'' \times 65'' $ and consists of 331 science fibres of $2.7''$ diameter packed in an hexagonal pattern, leading to a filling fraction of $65 \% $.

The data covers the spectral range between 3700 $\mathrm{\AA}$ - 7100 $\mathrm{\AA}$, with a spectral resolution of FWHM $\sim 8 \mathrm{\AA}$. 
Due to the large projected size of the galaxy ($10.5' \times 9.5' $) the data consists of a mosaic of observations taken on 6 different nights over a period of 3 years. The final size of the mosaic is of $6' \times 7'$. We estimate the PSF to have a FWHM of approximately $3''$. 

\subsection{H\textsc{I} 21 cm observations}
We used observations of the 21 cm line from THINGS (The H\textsc{i} Nearby Galaxy Survey, \citealt{Walter2008}). This survey comprises 34 nearby galaxies observed with the VLA, at high spectral ($< 5.3  \ \mathrm{km \ s^{-1}}$) and spatial resolution. NGC 628 was observed using the B, C and D configurations with a combined on-source integration time of $\sim 10$ hr. We use the publicly available\footnote{http://www.mpia.de/THINGS/Data.html} `robust' weighted map (the `robust' weighting scheme is used to achieve a sensitivity close to natural weighting while preserving a resolution close to uniform weighting, see \citealt{Walter2008} for details),  which has a beam size of $ \approx 6''$ ($\mathrm{B_{min}=5.57'', B_{maj}=6.8''}$). The $1 \sigma$ noise per 2.6 $\mathrm{km \ s^{-1} }$ channel is $\mathrm{0.66 \ mJy \ beam^{-1}}$, corresponding to a sensitivity of $\mathrm{ \approx 0.5 \ M_{\odot}  \ pc^{-2} }$, sufficient to trace the atomic hydrogen in all regions where it constitutes the dominant component of the ISM.

\subsection{CO(2 -1) observations}
Observations of the CO(2 -1) transition we used to infer the molecular gas content. Maps are obtained from the publicly
available\footnote{http://www.mpia-hd.mpg.de/HERACLES/Data.html} HERA CO-Extragalactic Survey (HERACLES,
\citealt{Leroy2009}). HERACLES is a survey of 48 nearby galaxies using the HERA multi-pixel receiver on the IRAM 30 m
telescope, with $13''$ angular resolution and $2.6 \ \mathrm{km \ s^{-1}}$ spectral resolution. The $1 \sigma$ sensitivity
of the map is estimated to be $\mathrm{ \approx 3 \ M_{\odot} \ pc^{-2}}$ (with the Galactic conversion factor discussed
in Sect.3).

\subsection{Multi-Wavelength photometric data}
Photometry in different bands was collected to obtain a reliable estimate of the stellar mass
surface density by performing pixel-by-pixel spectral energy distribution (SED) fitting. We collected observations from GALEX (Galaxy Evolution Explorer), 2MASS (Two Micron All Sky Survey) and the Spitzer Space Telescope IRAC camera. In the optical band, we make use of the photometry from SDSS Data Release 7. Table \ref{multiwav} summarises
the basic properties of the dataset used and the relevant references.

\begin{table*}
\caption{Multi-wavelength observations of NGC 628 used in this work to supplement the IFS mosaic from \protect\cite{Rosales-Ortega2010}.}
\label{multiwav}
\centering
 
\begin{tabular}{ l c c c c c c}

 \hline 
  Telescope 	& Filter/Receiver 	& $\mathrm{\lambda_{eff}}$ [\AA]  & Bandwidth [\AA]  & PSF FWHM $['']$ & Reference   \\
  \hline 
  \hline
  GALEX	 & FUV 	&  1516 			& 		268 		&	4.3		& \cite{Morrissey2007}	\\
  	 			&NUV	&	2267			&		732			&	5.3		& ''			 	\\
  \hline	
  SDSS	& $\mathrm{u'}$		&	3540			&		599			&	1.49		&  \cite{Abazajian2009}		 \\
   		& $\mathrm{g'}$		&	4770			&		1379			&	1.34		& ''				 \\
		& $\mathrm{r'}$		&	6222			&		1382			&	1.16		& ''				 \\
		& $\mathrm{i'}$		&	7632			&		1535			&	1.03		& ''				 \\
		& $\mathrm{z'}$		&	9060			&		1370			&	1.13		& ''				 \\
  \hline
  2MASS	& J		&	1.235 [$\mu m$]		&		0.162  [$\mu m$]			&	2.5		&  \cite{Skrutskie2006}	 \\
    		& H		&	1.662  [$\mu m$]		&		0.251  [$\mu m$]			&	2.5		& ''					 \\
		& $ \mathrm{K_s}$	&	2.159  [$\mu m$]		&		0.262  [$\mu m$]	&	2.5		& ''					 \\
		
  \hline
  Spitzer	& IRAC1		&	3.550 [$\mu m$]		&		0.75  [$\mu m$]			&	1.66		& \cite{Fazio2004}	 \\	
  		& IRAC2		&	4.493  [$\mu m$]		&		1.01  [$\mu m$]		&	1.72		& ''		 \\	
	\hline
   	\hline
   	IRAM 30m CO (2-1)	& HERA		&	$\nu = 230.54 \ \rm [GHz]	$ 	&		&	13		&  \cite{Leroy2009}		 \\	
   	\hline
   	VLA	H\textsc{i} &   	&	$\nu = 1420.40 \ \rm [MHz]	$	&			&	5.57 $\times$ 6.80		&  \cite{Walter2008} \\	
  
\end{tabular}
\end{table*}

\section{Data analysis}
\label{analysis}

In this section we present the analysis performed on the reduced data to obtain gas phase metallicities, stellar mass surface density and gas mass surface density profiles. We also briefly discuss the adopted choice of gas phase metallicity calibrators (Sec. \ref{gas_metall}) and the radial gradients of the derived quantities (Sec. \ref{grad2}). Throughout the section we highlight potential sources of systematic uncertainty.

\subsection{Emission line fluxes} \label{em_lines_flux}

We extract emission line fluxes in each spaxel from the IFS datacube following a very similar procedure to
\cite{Belfiore2015a}. Here we give a brief overview of the main steps involved.

\begin{enumerate}


\item{For each spaxel, we fit a linear combination of single stellar population (SSP) templates, after correction for
systemic velocity and instrumental dispersion, using Penalised Pixel Fitting \citep{Cappellari2004}. In this work we used
a grid of  29 SSP templates, generated by using \cite{Maraston2011} models based on the empirical STELIB spectral library\
\citep{leBorgne2003}. A spectral region of $600 \ \mathrm{ km \ s^{-1}} $ is masked around each emission line we wish to fit. Strong sky lines are also masked. We do not attempt to extract stellar population parameters from the SSP fits. Such a study has recently been carried out by \cite{Sanchez-Blazquez2014}, and in Sec. \ref{budg} we make use of the stellar metallicity gradient derived in their work.} 

\item{For each spaxel, we subtract the stellar population fit from the observed spectrum to obtain a pure emission line
spectrum. This is then fitted with a set of Gaussian functions, one per emission line (using least-squares minimization).
The doublets $\mathrm{[OIII]  \lambda \lambda 4959, 5007 \ and  \ [NII] \lambda \lambda 6548, 6584 }$ are set to have the
same velocity and velocity dispersion and the ratio of their intensities is fixed to the theoretical one. We calculate emission line fluxes by integrating the flux under the fitted Gaussians.}

\item{We apply a signal to noise cut to the emission line maps generated above, imposing a S/N larger than 5 in $\mathrm{H
\alpha \ and \ H \beta}$. This threshold is intentionally high to exclude regions of diffuse emission (mainly inter-arm
regions) where the $\rm H \alpha$ emission might not be ascribed to bona-fide H\textsc{ii} regions. We also discard regions where both the $\mathrm{[OII]}$  and the [OIII] line are undetected.}
\item{We calculate the reddening from the Balmer decrement, by using the $\mathrm{H \alpha  / H \beta}$ ratio and a
\cite{Calzetti2001a} attenuation curve with $\rm R_V = 4.05$. The theoretical value for the Balmer line ratio is taken from \cite{Osterbrock2006}, assuming case B recombination ($ \mathrm{H \alpha  / H \beta=2.87 }$). We note that the use of extinction curve of \cite{Cardelli1989} (or the modification by \citealt{O'Donnell1994}) with $\rm R_V=3.1$ yields very similar results for the 3600 $\rm \AA$ to 7000 $\rm \AA$ wavelength range considered in this work.}

\end{enumerate}

\subsection{Gas phase metallicity} \label{gas_metall}

Reliably measuring the gas metallicity from emission lines ratios remains a difficult problem in observational astrophysics. Different line ratios (`diagnostics') and methods (`calibrations') have been developed for the task. However it is well-known that different calibrations, even when based on the same diagnostics, can give results differing by up to 0.6 dex \citep{Kewley2008, Lopez-Sanchez2012,
Pena-Guerrero2012}. It is beyond the scope of this work to resolve the abundance scale problem (but see recent advances towards a possible solution in \citealt{Dopita2013, Perez-Montero2014, Blanc2015}). Fixing the gas phase metallicity scale is, however, a key factor in determining the total metal budget. We therefore discuss some of the assumptions going into widely used metallicity calibrators and possible shortcomings of each.

The metallicity (i.e. the oxygen abundance $\rm 12+log(O/H)$, where O and H are number densities),
can be calculated knowing the electron temperature, which can be derived from the ratio of the oxygen
auroral line $\mathrm{[OIII] \lambda 4363}$ and another oxygen line, like $\mathrm{[OIII] \lambda 5007}$. This method
(referred to as the $\mathrm{T_e}$ method\footnote{Or sometimes as the `direct method'. However, given the well-known
systematic uncertainties in $\rm T_e$-measured metallicities, we feel that this designation is inappropriate}) is generally
considered one of the most reliable, at least in the low metallicity regime \citep{Pagel1992, Garnett1992, Izotov2006}. At
high metallicities, however, the electron temperature decreases and the auroral line is generally undetected. Moreover,
temperature fluctuations make the $\rm T_e$ method unreliable at high metallicities \citep{Stasinska2005, Bresolin2007}. Due to
the strong temperature dependance of the emissivity of the auroral lines, regions with higher than average
temperature may dominate the line emission, hence biasing the metallicity measurements towards low values.
Finally, there is evidence that the distribution of electron energies in H\textsc{ii} regions might not follow
a Boltzmann distribution \citep{Binette2012, Nicholls2013}, hence making the definition of temperature arguable, while a $\kappa$-distribution may be more appropriate \citep{Dopita2013}.

Recombination lines can also be used to estimate the metallicity. These are almost insensitive to temperature, which
makes them much less dependent on temperature fluctuations and on the specific distribution of electron energies.
However, these lines are approximately $\rm 10^{4}$ times fainter than $\rm H \beta$, which makes their detection possible only for bright sources on 8-meter class telescopes. Metallicities estimated from the recombination lines are systematically offset, by about 0.2
dex, compared to those measured with the $\mathrm{T_e}$ method \citep{Blanc2015}. Some authors \citep{Peimbert2010, Pena-Guerrero2012} attribute this discrepancy to the presence of small temperature fluctuations within H\textsc{ii} regions, which (as
discussed above) would bias the observed auroral line fluxes towards the highest temperature, lower abundance regions.

Finally, several authors have calibrated strong nebular line ratios (making use of $\mathrm{[OIII] \lambda \lambda 4959, 5007,
\ [OII] \lambda 3727, \ [NII] \lambda 6584 }$, etc) as abundance diagnostics. While the intensity of these lines depends on
several factors besides the metal abundance (ionization parameter, density, N/O ratio), calibrations take advantage of the
fact that in the H\textsc{ii} regions observed in the local Universe relations exist between some of these parameters. Strong
line diagnostics can be calibrated against $\mathrm{T_e}$-based metallicity measurements \citep{Pilyugin2010, Pilyugin2005,
Pettini2004}, photoionisation models \citep{Denicolo2002, Tremonti2004, Kobulnicky2004}, or a mixture of both
\citep{Nagao2006, Maiolino2008}. Recently, \cite{Blanc2015} has shown that some (but not all) photoionisation model grids are able to reproduce the metallicities measured using recombination lines.

To take into account the systematic uncertainty introduced by the choice of metallicity calibration, in this work we use two independently derived calibrations. 
\begin{enumerate}
\item{The calibration from \cite{Maiolino2008} (M08), which is based on photoionisation models \citep{Kewley2002} in the high metallicity regime and anchored to $\rm T_e$-based metallicity measurements in the low metallicity regime. While M08 provides calibrations for various strong line diagnostics, here we use only the calibration based on the R23 parameter,
\[
\mathrm{R_{23} =  ([OII] \lambda 3727+ [OIII]\lambda 4959 + [OIII]\lambda 5007) / H \beta}.
\]
In the context of this work we have re-calculated the parametrisation of the R23-metallicity relation using the same
procedure in M08, but using a 5th order polynomial fit (M08 used a fourth order polynomial) to provide a better fit to the
high-metallicity end of the relation between 12+log(O/H) and R23. Overall the M08 calibration gives similar abundances to
those obtained using the calibration used by \cite{Tremonti2004} and the calibration of R23 presented by
\cite{Kobulnicky2004}. Since the R23 parameter is double-valued, we break the degeneracy by using the [NII]/[OII] ratio, concluding that all the regions in NGC 628 belong to the upper branch of R23. We note that the use of a calibration the recursively solves for both metallicity and the ionisation parameter \citep{Kobulnicky2004} does not substantially alter the derived metallicities.}
\item{The calibration from \cite{Pettini2004} (PP04), based mostly on $\rm T_e$ measurements, using the O3N2 index:
\[
\rm O3N2 = log \frac{[OIII] \lambda 5007 / H \beta }{[NII] \lambda 6584 / H \alpha}
\]
This calibration has the disadvantage of depending explicitly on the the nitrogen abundance, which does not necessarily
scale in a simple way with the oxygen abundance \citep{Perez-Montero2009, Perez-Montero2014, Belfiore2015a}, since nitrogen has both a primary and a secondary nucleosynthetic origin and is released into the ISM on longer timescales than oxygen. 

It is also worth noting that we are (perhaps unduly) extending the PP04 calibration to super-solar metallicities. In the
original work of \cite{Pettini2004}, only two H\textsc{ii} regions with super-solar metallicity and $\rm T_e$ measurements
are presented, while the high-metallicity end of the calibration is fixed by considering a small number of regions with
metallicities calculated using photoionisation models. Even in the most recent calibration of the O3N2 diagnostic,
presented by \cite{Marino2013}, only 5 H\textsc{ii} regions with super-solar metallicity are included, which makes any
attempt to `directly' calibrate the relation between O3N2 and metallicity in the super-solar regime subject to large extrapolation uncertainties.}
\end{enumerate}

\begin{figure*}
\includegraphics[width=0.49\textwidth,  trim=0 120 50 120, clip]{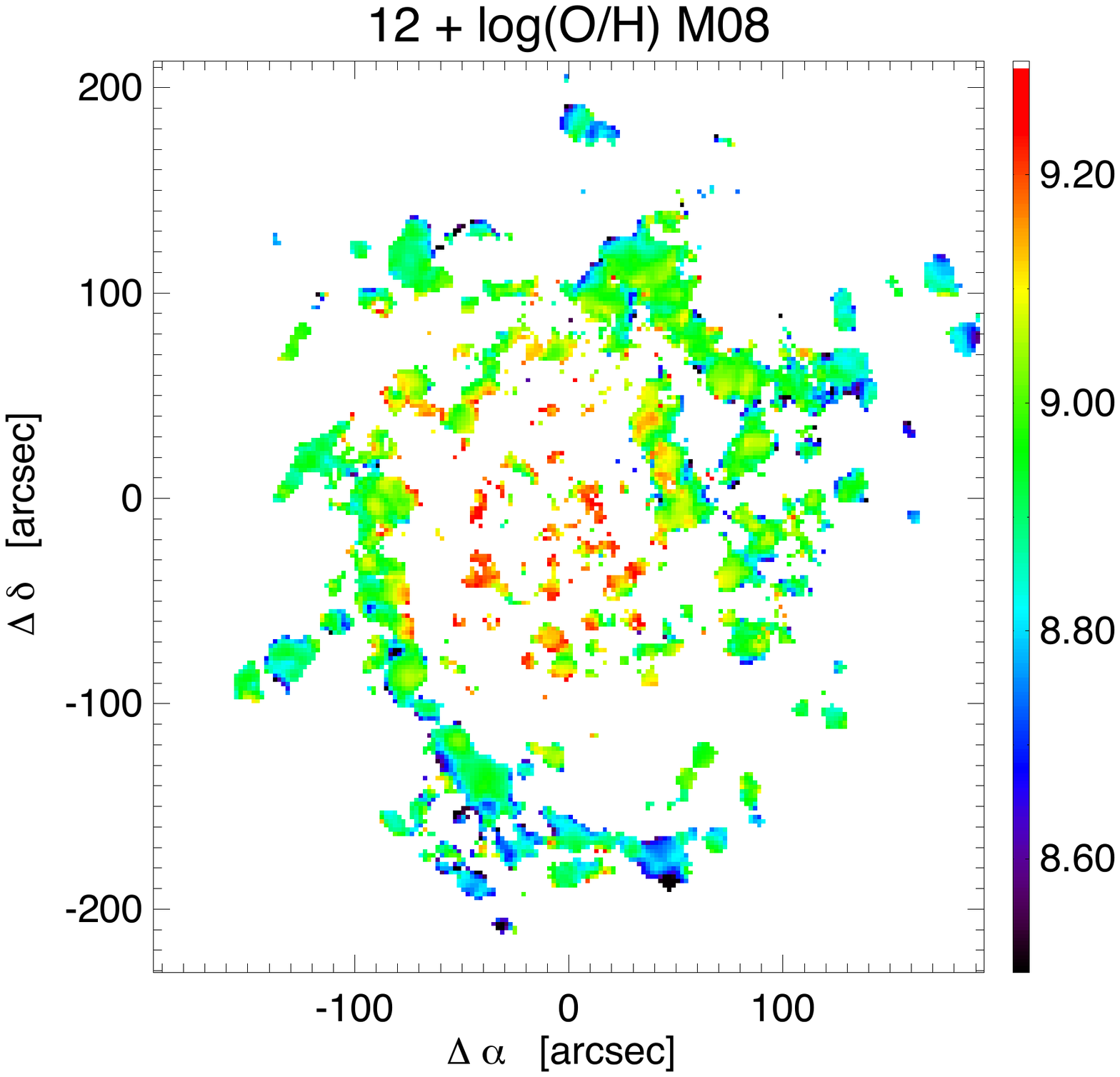} 
\includegraphics[width=0.49\textwidth,  trim=0 120 50 120, clip]{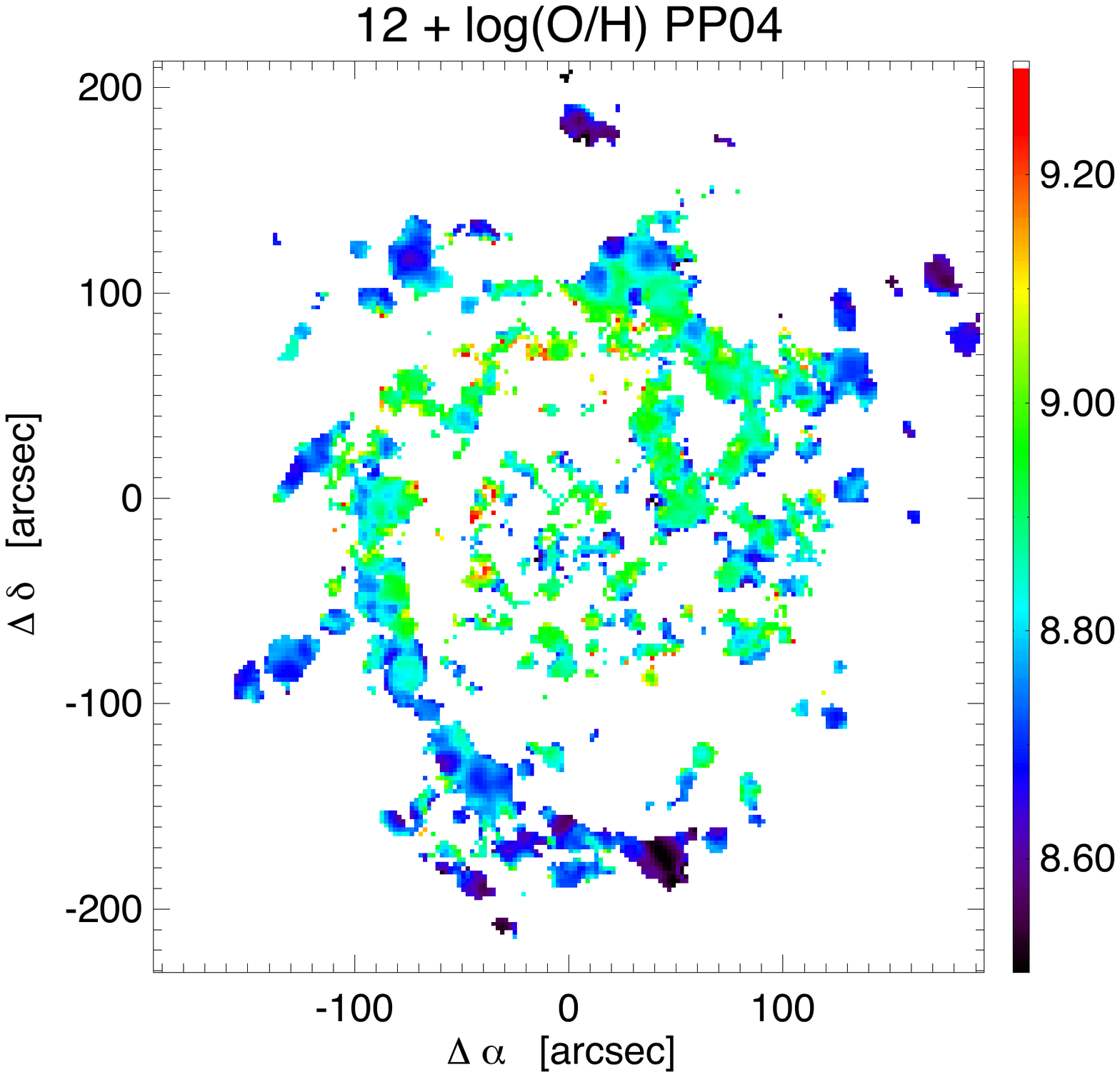}
\caption{Maps of the gas metallicity in NGC 628 using the \protect\cite{Maiolino2008} (M08) calibration based on R23 (left) and the
\protect\cite{Pettini2004} (PP04) calibration based on O3N2 (right). Only spaxels that lie below the \protect\cite{Kewley2001}
demarcation line in the $\rm [OIII]/H\beta$ versus $\rm [SII] /H \alpha$ BPT diagram are included.}
\label{metall_map}
\end{figure*}

We also note that the two adopted calibrations make different assumptions regarding dust depletion of oxygen. In the M08
calibration, an oxygen depletion factor of 0.22 dex is assumed \citep{Kewley2002}, while $\rm T_e$-based calibrations do
not correct for the amount of oxygen depleted onto dust. We follow \cite{Peimbert2010} and \cite{Pena-Guerrero2012} and
modify the metallicities obtained from both calibrations assuming a constant oxygen depletion factor of 0.1 dex. For the
rest of this work we will use gas phase metallicity to refer to the overall oxygen abundance of the ISM (gas and dust),
with the understanding that the dust is taken into account through a constant depletion factor. Altering the dust depletion by $\pm 0.1$ dex does not substantially change any of our conclusions.

Since M08 and PP04 are representative of the two main classes of metallicity calibrators ($\rm T_e$-based and photoionisation-model-based), alternative calibrations will not lead to metallicity estimates significantly higher than M08 or significantly lower than PP04. A thorough analysis of the gas phase metallicity in NGC 628 was presented in \cite{Rosales-Ortega2011}, who used a similar set of calibrators (as noted above, the \citealt{Kobulnicky2004} calibration is roughly equivalent to the adopted \citealt{Maiolino2008}) and obtained similar gradients and zero-points to the ones presented in this work. Interestingly, recent work from \cite{Croxall2013}, who used far-IR lines to estimate metallicity in NGC 628, points towards oxygen abundances within the range bracketed by the M08 and PP04 calibrations. Additionally \cite{Berg2015} obtained direct $\rm T_e$ measurements of metallicity for 45 H\textsc{ii} regions in NGC 628, which agree on average, though with large scatter, with the metallicities derived with the PP04 calibration. The reason for the large scatter in the $\rm T_e$-derived metallicities is not clear, but according to \cite{Berg2015} might be due to systematics in their electron temperature measurements.


Since metallicity diagnostics are only calibrated for star forming regions (classical H\textsc{ii} regions), regions of
galaxies where other types of ionisation (shocks, active galactic nuclei, evolved stars, etc) are dominant should be
excluded. We use the standard ionisation diagnostic diagram (BPT diagram \citealt{Baldwin1981, Veilleux1987, Kewley2001,
Kauffmann2003a}) to identify galactic regions as star-forming and discard regions that
lie above the \cite{Kewley2001} demarcation line in the $\mathrm{ [OIII]/H\beta}$ vs $\mathrm{ [SII]/H\alpha} $
diagnostic diagram. All of the discarded spaxels are found in the inter-arm regions and present LINER-like ionisation,
which might be due to evolved hot stars dominating the ionisation budget in regions where no recent star formation has
taken place \citep{Yan2012, Belfiore2015a}. We note that such regions, not complying with the H\textsc{ii} classification, are only
5\% of the total, and their exclusion does not affect our final results.

The resulting maps of gas phase metallicity for the M08 and PP04 calibrations are shown in Fig. \ref{metall_map}. For reference, the photospheric solar oxygen abundance, as derived by \cite{Asplund2009}, is $\rm 12+ log(O/H) = 8.69$, or equivalently $\rm Z_\odot (O)= 0.00585$. Note that we are referring to Z as the oxygen abundance by mass. The total metal content (fraction by mass of elements heavier than helium) in the Sun is $0.0142$.

\begin{figure*} 
\includegraphics[width=0.50\textwidth,  trim=0 90 50 180, clip]{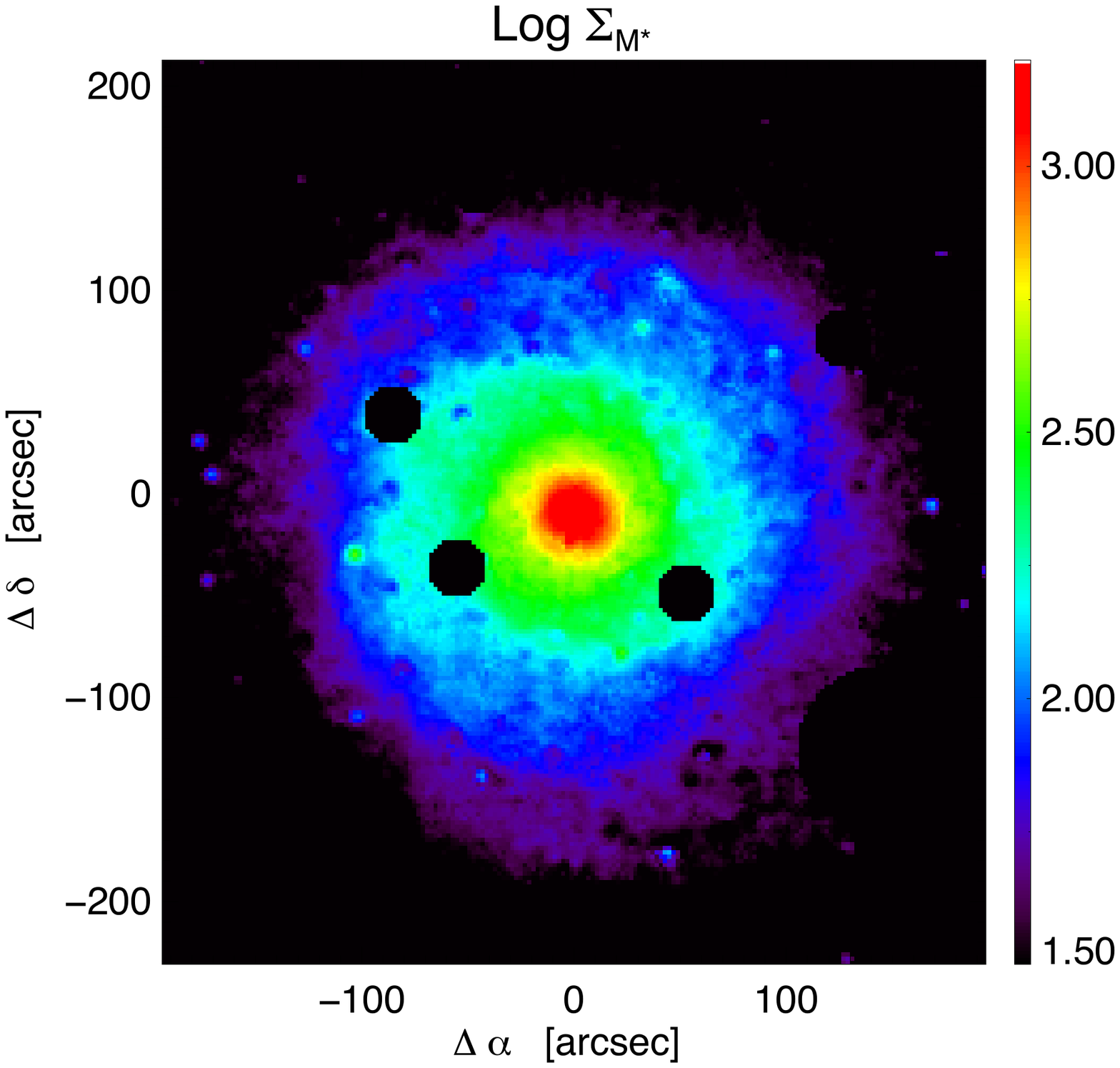} \quad
\includegraphics[width=0.47\textwidth, trim=45 90 60 120, clip]{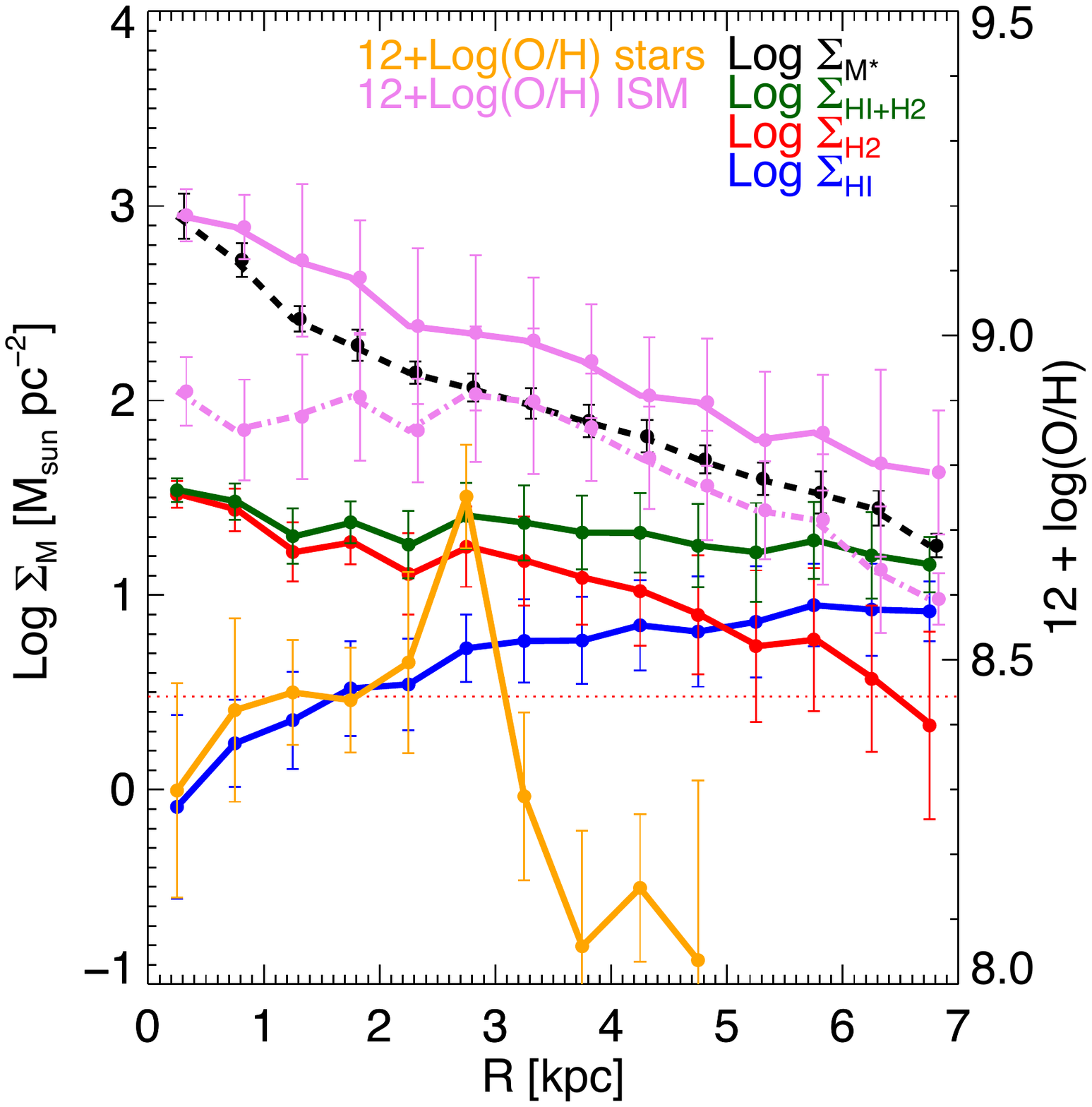}
\caption{Left: Stellar mass surface density map, created from the spatially resolved SED fitting. Foreground stars have been masked. Right: Radial
dependance of various physical quantities across the disk of NGC 628 (in radial bins; error bars represent the scatter in each radial bin). Atomic gas
surface density (blue) shows a dip in the central region, while the molecular gas surface density (red) is peaked at the centre and quickly decreases
outwards. The horizontal red dashed line corresponds to the sensitivity limit of the CO observations. The total gas content (green) is nearly flat across the whole disk up to 3 $R_{e}$ ($\rm 1 \ R_e = 2.4 kpc$). A clear radial gradient is observed in both stellar mass surface density (black, short dashes) and gas metallicity (violet). The latter is shown both for the M08 (solid violet) and PP04 (dot-dashed violet) metallicity calibrations. The stellar metallicity gradient from \protect\cite{Sanchez-Blazquez2014} is shown in orange. Note that the metallicities of gas and stars are plotted with respect to the alternative y-axis on the right in units of $\rm 12+ log(O/H)$.}
\label{grad}
\end{figure*}

\subsection{Stellar mass surface density}
Stellar mass surface densities are estimated by performing SED fitting in each pixel on the multi-wavelength photometric data from UV to near IR, using the software package \textsc{cigale} (Code Investigating GAlaxy Emission, \citealt{Burgarella2005, Noll2009}).
In detail, we followed the procedure outlined below (similar to \citealt{Boquien2012}).
\begin{enumerate}
\item{The different observations were smoothed to the lowest resolution of the photometric dataset (GALEX NUV), using the kernels from \cite{Aniano2011}.}
\item{The maps were then regridded on the same pixel scale as the IFS mosaic.}
\item{For each pixel, the flux and the flux error in each band was extracted. Errors are computed by taking into account the uncertainty in absolute calibration of the different instruments. Fluxes and errors are used as input for the software package \textsc{cigale}.}
\item{\textsc{cigale} creates FUV-to-FIR SEDs consisting of dust-attenuated complex stellar population models. In this analysis we used \cite{Maraston2005} templates (with a Kroupa IMF) to build a grid of stellar models with a range in metallicity (from 0.5 to 2 solar). The star formation history is parametrised by an exponentially decreasing SFR, and an old stellar population with and a grid of younger stellar population bursts. Extinction is fitted using a Calzetti-like attenuation curve. The optimal galaxy parameters are derived by \textsc{cigale} using a Bayesian-like analysis. A map of stellar mass surface density is then created (Fig. \ref{grad}, left panel)}
\end{enumerate}
The formal errors computed by \textsc{cigale} (which take into account the uncertainty in the data, but not the systematics of the method) are of the order of 0.1 dex. As a further check, we compared the stellar mass estimated by \textsc{cigale} with the stellar mass  estimate obtained using the procedure in \cite{Bell2003} and using the u-r colour to estimate the mass-to-light ratio. The stellar mass densities obtained with the Bell formulation are consistent with the ones obtained with \textsc{cigale}, with a small systematic difference of less than 0.1 dex at very large radii ($\rm R > 8 \ kpc$). The total mass of NGC 628 out to $\rm 3 \ R_e$ computed by \textsc{cigale} is $\rm \log(M_\star/M_\odot)=9.8$.

\subsection{Stellar metallicity} 

To take into account the metals locked in stars we make use of the recent study of \cite{Sanchez-Blazquez2014}, who present the mass-weighted stellar metallicity gradient for NGC 628 by performing full spectral fitting using the software package \textsc{steckmap} on the PINGS IFS data.
Despite the increased level of sophistication of the spectral fitting algorithms in recent years \citep{Heavens2000, Panter2004, CidFernandes2005, Gallazzi2006, Sanchez-Blazquez2011, CidFernandes2014, Wilkinson2015, McDermid2015a}, stellar metallicity remains the hardest quantity to estimate in stellar population studies. The difficulty is due to the non-linear effect of a limited set of metallicity values generally available in spectral libraries (since most stellar libraries rely on observations in the solar neighbourhood super-solar metallicity stars are not well represented) and the well-known age-metallicity degeneracy \citep[e.g.][]{Worthey1994}, which acts in the direction of confusing old, metal poor systems with young metal rich ones. While the ability of spectral fitting codes to reliably derive input parameters increases as a function of S/N, it has been shown (see for example \citealt{CidFernandes2014, Wilkinson2015}) that even for high S/N spectra, stellar metallicity calculated with different sets of simple stellar populations can differ by more than 0.2 dex. It is beyond the scope of this work to perform a new full spectral fitting study of the PINGS data for NGC 628 and further evaluate these systematic uncertainties. We note, however, that our simple approach of propagating of the errors in the stellar metallicities presented in \cite{Sanchez-Blazquez2014} (median error $\sim 0.15$ dex) might represent an underestimate of the real uncertainty.

Taking into account the necessary correction factor due to the fact that \cite{Sanchez-Blazquez2014} adopt a different distance for NGC 628, we can obtain a stellar metallicity gradient out to 5 kpc. We put the gas phase and stellar metallicity on the same abundance scale by using the solar metallicity of \cite{Asplund2009} and assuming solar abundance ratios. We do not attempt to correct for possible non-solar abundance ratios (e.g. $\alpha$-enhancement) in the bulge, since in any case the correction will only affect our innermost radial bin and will not have a significant effect on the subsequent analysis of integrated properties.

\subsection{Gas mass surface density}
The H\textsc{i} mass can be calculated directly from the H\textsc{i} surface brightness, assuming that H\textsc{i} is everywhere optically thin. 

%
%

To convert from CO luminosity to $\mathrm{H_2}$ mass we adopt a constant CO to $\mathrm{H_2}$ ($\alpha_{CO}$) conversion factor given by

$$\mathrm{  \alpha_{CO}= 4.35 \frac{M_\odot \ pc^{-2}}{K \ km \ s^{-1}} 		}$$
This value of the conversion factor is appropriate for the Milky Way \citep{Solomon1987, Strong1996, Abdo2010, Bolatto2013}. We explored the effect of using a metallicity-dependant conversion factor \citep{Schruba2011} and the conversion factor derived by \cite{Blanc2013} by inverting the star-formation law in NGC 628, finding that none of our conclusions are substantially modified.
As the conversion factors are calibrated for the CO (1-0) line, while the HERACLES observations map the CO (2-1) transition, we assume a constant ratio of 0.7 between the two CO luminosities \citep{Leroy2012}.

To calculate the total gas mass we multiply the H\textsc{i} component by 1.36 to account for helium and add the molecular gas mass computed using the conversion factor above, which already takes helium into account.

\subsection{Radial gradients} 
\label{grad2}

It is well known that spiral galaxies exhibit radial gradients in their physical properties, including stellar mass surface density and metallicity \citep{Bell2001, Moustakas2010, Delgado2013, Sanchez2014}. 

In Fig. \ref{grad} we present the radial gradients for the different physical quantities that we have derived in this section for NGC
628. We observe an exponential profile in stellar mass surface density and gas phase metallicity (black dashed and violet
lines respectively). 

We note here that the PP04 gas metallicity gradient presents a flattening in the inner 2.5 kpc ($\rm \approx 1 \ R_e$, violet dot-dashed line), already reported in previous work \citep{Rosales-Ortega2011} and consistent with the flattening observed in a subsample of galaxies from the CALIFA survey \citep{Sanchez2014}. With the current data it is not possible to assess whether this feature corresponds to a real abundance drop in the central region or if it is an artefact of the metallicity calibration. In particular, we note that such flattening observed when using the O3N2 calibration is often associated with a central decrease of the N/O abundance ratio (which can be traced by the [NII]/[OII] flux ratio), and this is also the case of NGC 628. Therefore the metallicity flattening may actually reflect a relative abundance variation affecting the metallicity diagnostic used in PP04, rather than a global metallicity flattening. 
However, we note that from the theoretical standpoint, bars and other non-axisymmetric perturbations can trigger large scale gas flows and induce a flattening of the metallicity gradient \citep{Roskar2008a, Roskar2008b, Cavichia2014}, together with a metallicity `hump' at the interface. While the gas phase abundance does not show evidence for this hump, a sharp increase in the stellar metallicity is reported at 2.5 kpc by \cite{Sanchez-Blazquez2014} (orange line in Fig. \ref{grad}). They interpret this feature as the effect of an oval distortion (possibly a signature of a dissolving bar). The presence of a circum-nuclear ring of enhanced star formation seems to confirm this hypothesis.

The H\textsc{i} disc is much more extended than the FoV of the IFS data and presents clumpy substructure (seen in Fig. \ref{pic1}). The H\textsc{i} surface density profile is characterised by a central dip and and an increase towards larger radii. The maximum in H\textsc{i} surface density coincides roughly with the edge of the IFS field of view ($\rm  \approx 3 \ R_e$), hence in our subsequent analysis, which is limited to the radial extent of the IFS observations, we neglect about $40\% $ of the mass of the H\textsc{i} disc. We will discuss the importance of this extended H\textsc{i} disc in Sec. \ref{dis}.

The molecular gas profile is centrally concentrated and its surface density decreases with radius. The sensitivity limit of the HERACLES data is shown as a horizontal dotted line in Fig. \ref{grad}. The combination of the atomic and molecular gas profile generates a total gas surface density profile (green line in Fig. \ref{grad}) which is remarkably constant over the whole field of view of the IFS data. This implies that the gas fraction ($\rm f_{gas} = M_{gas}/(M_\star + M_{gas}) $) increases outwards roughly log-linearly with radius.

\section{The chemical evolution framework} \label{chem_evo}

In this section we present the analytical framework we will be using to study the metallicity evolution in resolved regions of NGC 628. 

We make use of simple relations obtained by invoking the instantaneous recycling approximation and perfect instantaneous
mixing. This framework has been proven successful in modelling the chemical evolution of oxygen \citep{Zahid2012, Dayal2013, Peeples2014, Peng2014}, which is predominantly produced by short-lived massive stars dying as core-collapse supernovae. We note in passing that the chemical abundances of other common elements (like N or Fe) require a more detailed model, taking into account stellar lifetimes, the Type Ia supernovae delay time distribution and variation of the nucleosynthetic yields with metallicity, and will not be attempted in this work. However, since oxygen is the most abundant metal by mass, it is a good tracer for the total metal content. 

Several studies in the literature have shown that, in order to interpret observed chemical abundances in galaxies, it is
necessary to devise a model taking gas inflows and outflows into account (often dubbed the `gas regulatory' or `bathtub' models; 
\citealt{Matteucci1986, Gibson2003, Lilly2013, Dekel2014, Peng2014}).

Within the instantaneous recycling approximation, denoting the oxygen fraction (by mass) in the ISM as $\rm Z_g$, the gas
mass as $\rm M_g$, the star formation rate as $\rm SFR$, the stellar mass as $\rm M_\star$, the total mass in oxygen in the galaxy as $\rm M_Z$, the mass of oxygen in the ISM as $\rm M_{Zg}$, the outflow rate as $\rm \Psi$ and the inflow rate as $\Phi$, the galaxy's chemical evolution in a bathtub model is described by the following constitutive equations

\begin{equation}
\rm \frac{dM_g}{dt} = \Phi - (1-R) \ SFR - \Psi, \label{eqa1}
\end{equation}

\begin{equation}
\rm \frac{dM_\star}{dt} = (1-R) \ SFR, \label{eqa2}
\end{equation}

\begin{equation}
\rm \frac{ dM_{Zg} }{dt} = \frac{d (M_g ~Z_g)}{dt} = p \ SFR - \Psi Z_g - (1-R) \ Z_g \ SFR, \label{eqa3}
\end{equation}
 
and
\begin{equation}
\rm \frac{ dM_Z }{dt} = p \ SFR - \Psi Z_g, \label{eqa4}
\end{equation}

where p is the mass of newly synthesised oxygen per unit mass of gas converted into stars (we will refer to it as the \textit{yield} per stellar
generation) and R is the fraction of gas mass that is promptly returned to the ISM via stellar mass loss processes (referred to as the \textit{return fraction}). We emphasise that the same relations apply also to the total abundance of metals (after replacing the appropriate value for the net nucleosynthetic yield), if the abundance ratios of different elements are assumed constant.

Both the average yield and the return fraction are a function of the initial mass function (IMF) and in general also depend on metallicity and time. In our simple model neither the time nor the metallicity dependance of R and y will be further considered, since the oxygen yield is shown to be approximately independent of metallicity and the effect of stellar lifetimes on the return fraction can be considered negligible for our purposes
\citep{Thomas1998, Kobayashi2006, Zahid2012, Vincenzo2015a}. 

If we denote as p(M) the net yield of oxygen for stars of mass M, then the net yield (p) per stellar generation is given by
\begin{equation}
\rm p = \frac{\int_{M_{long-liv}} ^{M_{up}} p(M) \ IMF(M) \ dM}{\int_{M_{low}} ^{M_{up}} IMF(M) \ dM},
\label{yield}
\end{equation}
where the $\rm IMF(M) $ is the initial mass function, $\rm M_{low}$ and $\rm M_{up}$ are the lower and upper mass cutoffs of the IMF,  while $\rm M_{log-liv}$ is the highest mass of the stars that are considered `eternal' or, equivalently, the lowest mass of the stars contributing to the chemical enrichment. Note that an alternative definition of yield is given by the new mass of oxygen produced per unit mass \textit{of long-lived stars \citep{Searle1972} (in this case generally denoted as y). These two definitions are trivially related by the return fraction with $y = p/(1-R)$.\footnote{Note that \cite{Peng2014} does not follow this traditional notation. The quantity denoted as y in \cite{Peng2014} is the same quantity which we denote as p in this work.}}
It is important to stress that the yield per stellar generation is strongly dependent on the IMF, and not only on the `stellar' nucleosynthetic yield p(M). In fact, the average yield can vary by a factor larger than three for the same choice of stellar yields if the IMF is changed from \cite{Kroupa1993} to \cite{Chabrier2003} \citep{Vincenzo2015a}.

In this work, when a numerical value for the average yield is required, we make use of the results in \cite{Vincenzo2015a}, based on the compilation of stellar yields in \cite{Romano2010}, which have been shown to successfully reproduce the oxygen abundance in the Milky Way. We adopt a KTG IMF \citep{Kroupa1993}, $\rm M_{long-liv} = 1 \ M_\odot$ and therefore get a fiducial average yield of $\rm p = p_O= 0.007$ for oxygen and $\rm R = 0.30$ \citep{Vincenzo2015a}.
The same results can be applied to the {\it total} content of metals, by replacing the numerical value of the average yield with $\rm p = p_Z(tot) =0.013$.

It can be easily shown that in the case of a `closed box' evolution, i.e. no inflow and no outflow ($\rm \Phi=0$ and $\rm \Psi=0$), the gas metallicity is given by
\begin{equation}
\rm Z_g = \frac{p}{1-R} \ln{(f_{gas}^{-1}) } = y \ln{(f_{gas}^{-1}) },
\label{closed_box}
\end{equation}

where $\rm f_{gas} \equiv M_g/(M_g+M_\star)$ is the gas fraction. Some authors find convenient to define the so-called `effective yield' as
\begin{equation}
\rm y_{eff} \equiv \frac{Z_g}{ \ln{(f_{gas}^{-1}) } }.
\label{eq_yeff}
\end{equation}

\begin{figure}
\centering
\includegraphics[width=0.47\textwidth, trim=50 100 30 130, clip]{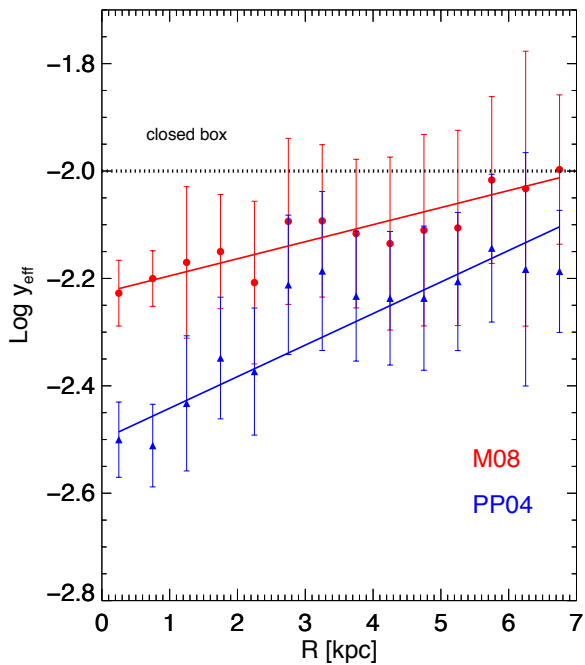}
\caption{The radial variation of the effective yield in NGC 628. The value expected for the closed box case,  $\rm y/(1-R) = 0.010$, is indicated as a
black dotted line. The red circles and blue triangles represent respectively the results obtained with the \protect\cite{Maiolino2008} and \protect\cite{Pettini2004} calibrations.}
\label{fig_yeff}
\end{figure}

Whenever $\rm y_{eff} \ne y = p/(1-R)$ Eq.\ref{closed_box} implies that the system has not evolved as a closed box. Fig.~\ref{fig_yeff} shows the radial gradient of the effective yield in NGC 628, clearly demonstrating that $\rm y_{eff}$ increases with radius. Importantly, the observed effective yield is always lower than $\rm y = p/(1-R)=0.010$, implying that the region of the disc of NGC 628 sampled by us has not been evolving as a closed box. 

However, the information provided by the effective yield is highly degenerate. More specifically, both enriched gas outflows or pristine (or low metallicity) gas inflows will lead to the observed effective yield to be lower than the closed box value \citep{Edmunds1990}.

\section{Results}
\label{results}

\subsection{A spatially resolved oxygen mass budget} \label{budg}

The aim of this section is to compute the total mass of oxygen present in NGC 628 as a function of radius and compare it with the total mass of oxygen expected to have been produced by its stellar component. 

This comparison provides a direct (model-independent) measurement of the mass of metals lost by each region in the galaxy via metal-enriched outflows. Moreover, the results obtained in this section can be directly compared with predictions from cosmological simulations.

\begin{figure*} 
\centering
\includegraphics[width=0.54\textwidth, trim=50 100 30 130, clip]{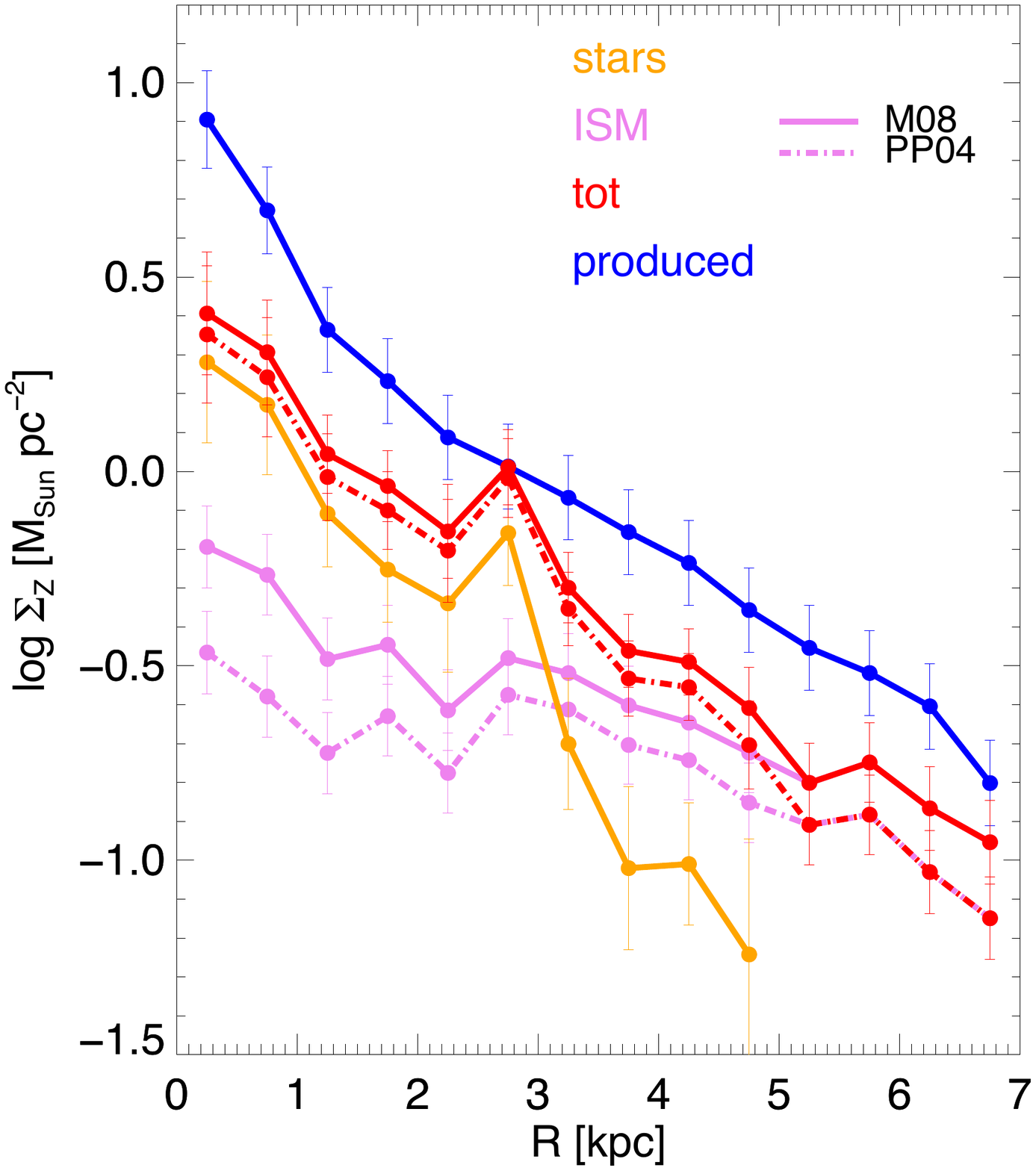}
\includegraphics[width=0.44\textwidth, trim=50 20 30 60, clip]{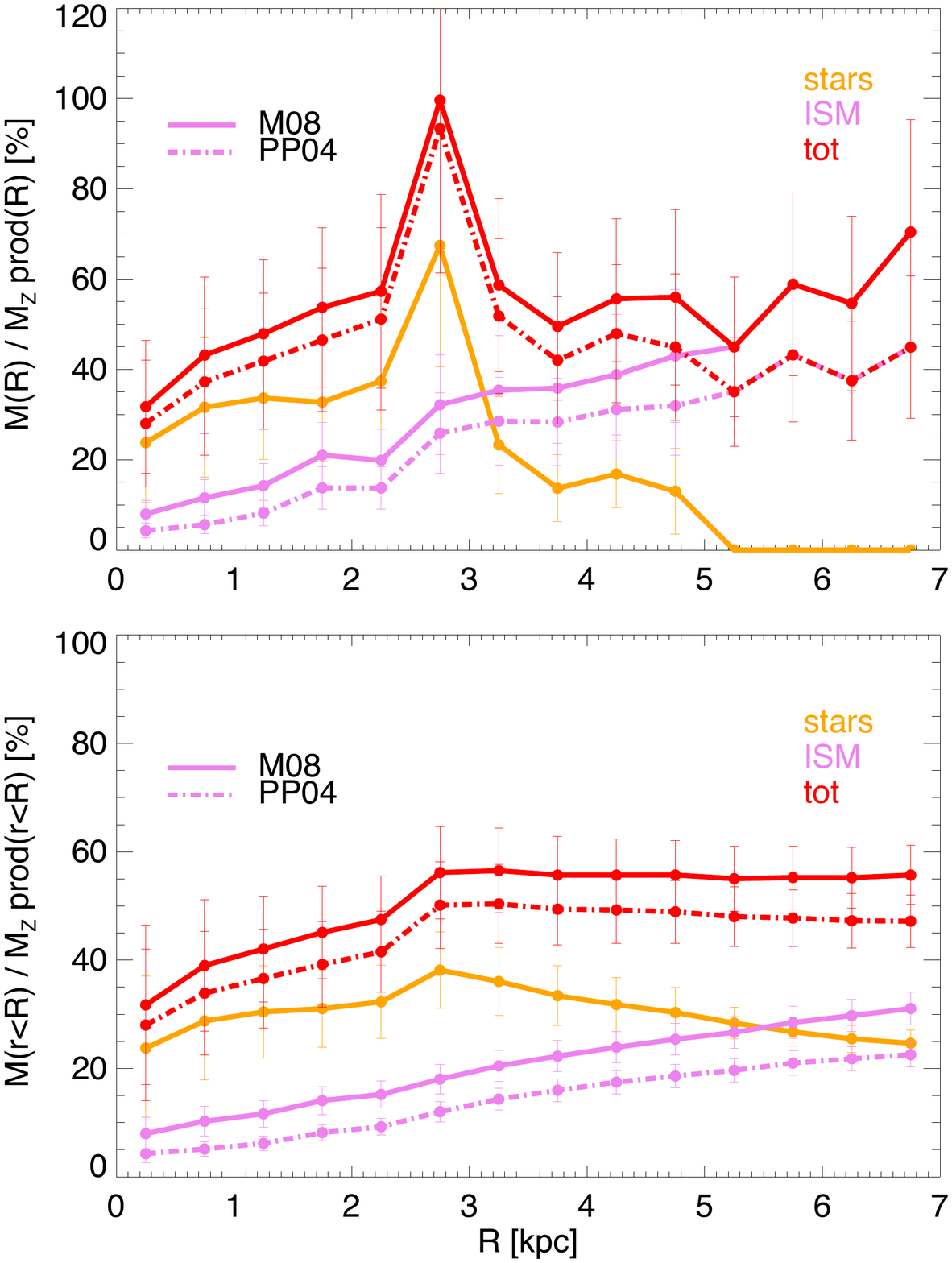}

\caption{Left: Oxygen mass per unit area in different galactic components compared with the oxygen mass expected to have been produced by the observed
stars, in different radial bins. The orange line shows the oxygen mass locked in stars, by using the stellar metallicity gradient from
\protect\cite{Sanchez-Blazquez2014}. The violet solid line shows the oxygen mass in the ISM obtained by using the M08 calibration, while the violet-dashed
line shows the oxygen mass in the ISM obtained by using the PP04 calibration. In red we show the total observed mass of oxygen (ISM + stars). The blue line
shows the mass of oxygen produced by the stars at the same location assuming our fiducial average yield.
Top Right: Mass of oxygen in each radial bin as a percentage of the mass of oxygen produced in the same radial bin. The colour-coding is the same as in
the left panel.
Bottom Right: The cumulative mass of oxygen within radius R in the different galactic component, expressed as a percentage of the total mass of oxygen
produced within the same radius R. The points shown for the last radial bin correspond to the integrated oxygen budget for NGC 628, showing that the galaxy as a whole has retained only $\approx 50 \%$ of the metals it has produced.
}
\label{met_budget}
\end{figure*} 

Following Eq. \ref{eqa4}, the total mass of oxygen \textit{produced} is given by

\begin{equation}
\rm M_{Z_{prod}} =  y \ M_{\star}. \label{eqd1}
\end{equation}

This corresponds to the total mass of oxygen actually present in the galaxy in the case of no outflows (Eq.~\ref{eqa4} with $\Psi =0$).
The total observed mass of oxygen is given by the sum of the mass of oxygen in the ISM ($\rm M_{Zg}= Z_{g} M_{g}$) and the mass of oxygen locked in stars
($\rm M_{Z \star}= Z_{\star} M_{\star}$), i.e.
\begin{equation}
\rm M_{Z(obs,tot)} = M_{Zg}+ M_{Z \star} = Z_{g} M_{g} + Z_{\star} M_{\star} ~.
\end{equation}

Both simulations \citep{Marinacci2014} and observations \citep{Shull2012, Tumlinson2013} argue that a large quantity of baryons and metals might reside in the hot haloes around galaxies. However, for the purposes of chemical evolution this gas is irrelevant since we assume it does not form stars, at least as long as it remains in the hot phase. Therefore in this work we consider this gaseous component as effectively external to the galaxy and the metals present in it as lost.

In Fig. \ref{met_budget} (left panel) we show the total observed mass of oxygen per unit area in stars (orange), in the ISM (violet), in ISM+stars (red) and the mass of metals that must have been produced by the observed stellar mass using our adopted yield (using Eq. \ref{eqd1}, blue line) as a function of radius. In the case of the ISM and ISM+stars lines, we show the results obtained with both the M08 (solid) and PP04 (dashed) metallicity calibrations. The error bars represent the error in the mean in each annulus, added in quadrature to the median of the intrinsic error in each of the physical quantities. The intrinsic error budget is dominated by the error on the stellar metallicity, followed by the uncertainty in metallicity and stellar mass (for both we get a median error of about 0.1 dex). These errors do not attempt to incorporate systematic effects, which would dominate the uncertainty budget for this study. Moreover we take the net yield (y) and the return fraction (R) to be known exactly. With these caveats, Fig. \ref{met_budget} demonstrates that using the fiducial yield \textit{at all radii more metals have been produced than can be accounted for considering both the metals in the ISM and in stars}.

In the discussion (Sec. \ref{dis}) we argue that a different choice of IMF (adopting for example Salpeter or Chabrier IMF) only makes the metal deficit worse. We also note that the mass of metals in stars presents a sharp feature between 2.5 and 3 kpc, which is a direct consequence of the sharp feature in the stellar metallicity gradient. Since it is present in only one radial bin, the effect of this feature on the overall metal budget of the galaxy is negligible and will not be discussed further.

In Fig.~\ref{met_budget}, top right panel, we show the mass of oxygen in stars (orange) and ISM (violet) as a function of
radius as a percentage of the total mass of oxygen produced in that annulus. The total mass of oxygen in each annulus (ISM + stars, red line) is also shown, and we plot the results using both the M08 (solid line) and PP04 (dot-dashed line) metallicity
calibrations. 

Fig.~\ref{met_budget} (top-right) shows that the fraction of metals lost is about 45\%-55\%, and approximatively constant within the uncertainties for $\rm R > 3 \ kpc$. There is also some evidence that the fraction of metals lost in the central region is higher (only $\sim$30\% of the metals have been retained). This feature may trace the effect of massive outflows during the formation of the bulge or the cumulative effect of nuclear activity in the past. Interestingly Fig.~\ref{met_budget} (top-right) also shows that the fraction of metals retained increases at R $>$ 5 kpc. Gas phase metallicity measurements at even larger radii provided by the \cite{Berg2015} dataset confirm and further extend this trend (see discussion in Sec. \ref{large_dist}).

Fig. \ref{met_budget} (lower-right) shows the \textit{cumulative} version of the previous plot. For each radius R, the mass of
oxygen in different components \textit{within} radius R is computed as a percentage of the mass of metals produced within the
considered radius. This plot can be interpreted as the results that would be obtained for the integrated properties of NGC
628 if all quantities (metallicities, gas masses etc) were measured using an aperture corresponding to a radius R on the sky.

We note that cumulatively the mass of oxygen in the ISM is sub-dominant out to $\rm R \approx 5.5 \ kpc$, at the latter
radius stars and ISM contribute roughly equally to the total oxygen budget. For easy reference, the masses of oxygen in
the different components within 7.0 kpc are summarised in Table \ref{table3}. Taking into account the uncertainty associated
with the choice of metallicity calibration, we conclude that out to R = 7.0 kpc the bulge and the disc of NGC 628 as a whole have lost $ \sim 50 \%$ of the oxygen they have produced. 

\begin{table}

\caption{The total oxygen mass in different galactic components, integrating the radial gradients for NGC 628 out to R = 7.00 kpc.}
\label{table3}
\centering
 
\begin{tabular}{ c c c }

\hline 
Metals	& 	Mass [$ \rm10^{7} \ M_{\odot}$] &  Mass [$\rm \% \ M_{z} \ prod$]   \\
\hline 
\hline
Stars		& 	$2.6 \pm 0.9$			& $25\pm 2 \%$ \\
ISM (M08) & 	$3.2 \pm 0.9$			& $31 \pm 3 \%$   \\
ISM (PP04) & 	$2.3 \pm 0.7$ 			& $23 \pm 2 \%$	\\
Stars + ISM (M08) &	$5.8 \pm 1.7$			& $56 \pm 5 \%$ \\
Stars + ISM (PP04) &	$ 4.9 \pm 1.5$			& $47 \pm 5 \%$ \\
\hline

\end{tabular}

\end{table}

\subsection{Inferred net outflow loading factor} \label{loading}

\begin{figure} 
\centering
\includegraphics[width=0.48\textwidth, trim=60 170 120 220, clip]{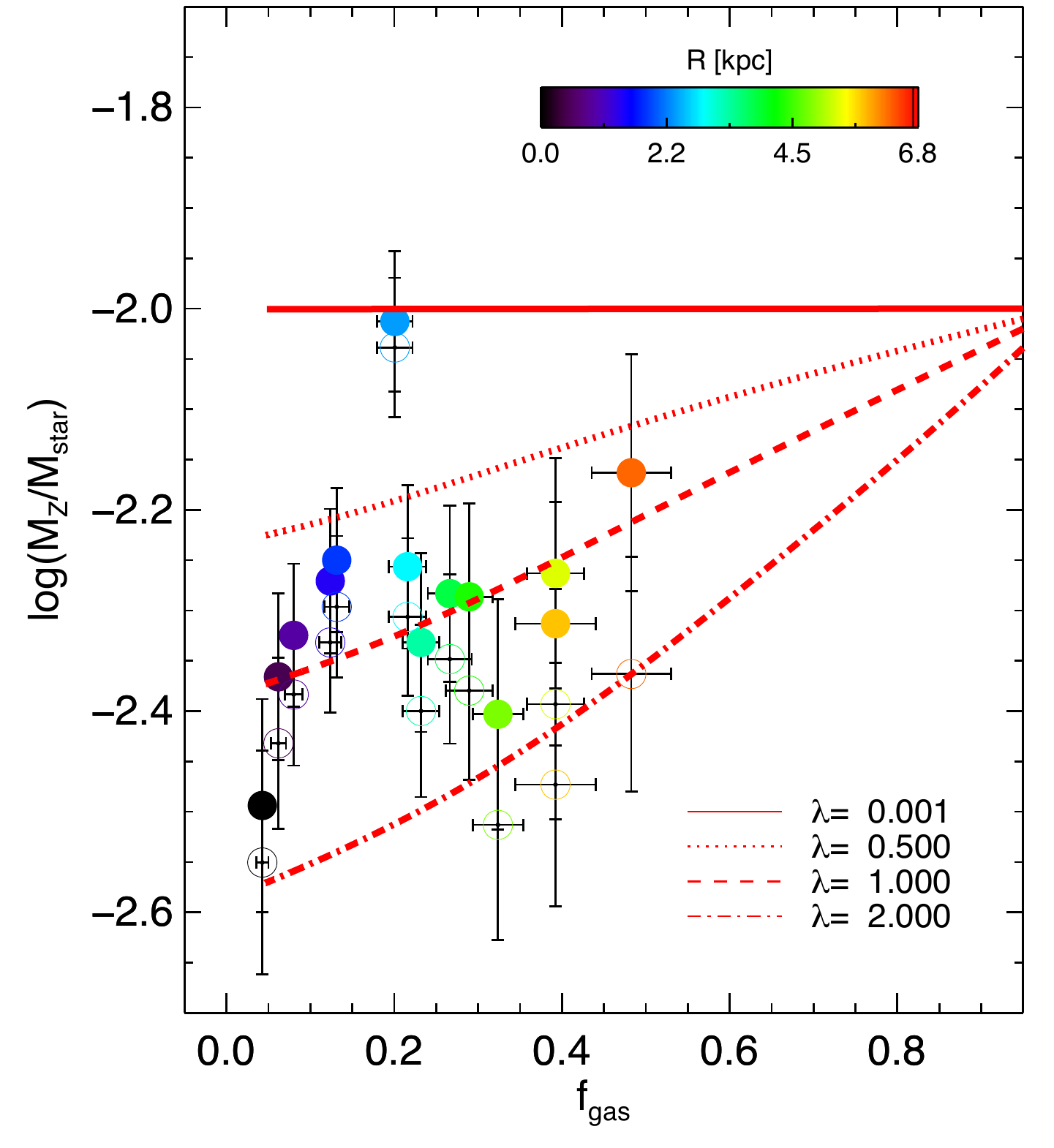}
\caption{Metals-to-stellar mass ratio ($\rm M_Z/M_\star$) versus gas fraction ($\rm f_{gas}$). Symbols show the values observed in radial annuli in NGC628,
where the colour-coding gives the galactocentric radial distance, as indicated in the colour bar. Solid symbols correspond to the mass of metals
in which the ISM metallicity is calculated by using the M08 calibration, while open symbols use the PP04 calibration. The red lines correspond to the prediction of simple chemical evolution models with different values for the average outflow loading factor $\lambda$. Note that the model relations are independent of the inflow rate $\Phi$ and of the star formation efficiency $\varepsilon$.}
\label{model_Mztot}
\end{figure} 

As discussed above, in the absence of outflows the total amount of metals produced should be simply given by the stellar mass times the amount of metals produced per mass of long lived stars (Eq.\ref{eqd1}). To model the effect of outflows we follow the formalism laid out in \cite{Peng2014} and assume the outflow rate $\Psi$ to be proportional to the star formation rate through a constant outflow loading factor $\lambda$,
\begin{equation}
\rm \Psi = \lambda ~SFR.
\end{equation}
The resulting total amount of metals is given by the equation
\begin{equation}
\rm M_{Z(tot)} =  y \ M_{\star} - \lambda \int Z_g(t)~SFR(t)~dt .\label{eqd1b}
\end{equation}

The last term is the `metal deficit' and its evaluation requires knowledge of the evolution of the metallicity with the star formation rate. To make further progress we therefore need to solve Eq. \ref{eqa1}--\ref{eqa2}. A general solution as a function of time can be obtained assuming a star formation law. Following previous work we use a simple linear relation of the form
\begin{equation}
\rm SFR = \varepsilon ~ M_{gas}
\end{equation}
where $\varepsilon$ is often refereed to as the \textit{star formation efficiency} or also as the inverse of the gas depletion time ($\rm 1/\tau_d$). 

The final piece of information needed is the functional form of gas inflow rate $\Phi$. A number of authors have assumed a gas inflow rate proportional to the SFR \citep[e.g.][]{Matteucci1983, Erb2008,Mannucci2009,Dayal2013, Troncoso2014,Kudritzki2015}. This assumption leads to simple analytical solutions of the resulting differential equations. Moreover it does not require consideration of the time variable, since all the quantities evolve in time like the SFR and hence the time dependance factors out. However, the assumption of an inflow rate proportional to the SFR has no physical grounds and has been used in previous work entirely for convenience. A much more realistic assumption is that of a constant, or slowly varying, inflow rate. As discussed in \citet{Peng2014} this assumption is much closer to the expectation based on the inflow rate of baryons in dark halos inferred in analytical calculations and numerical simulations of structure formation \citep{Faucher2011, Dekel2013}. 

Interestingly, analytical solutions of Equations \ref{eqa1}--\ref{eqa4} can still be obtained with the assumption of constant inflow rate, although they do depend explicitly on time. Exact solutions for $\rm M_{g}$ and $\rm M_{\star}$ are given in \citet{Peng2014}. Further exact solutions for $\rm Z_{g}$ and $\rm M_Z/M_\star$ are presented in Appendix \ref{app_A} of this work. Here, we exploit those solutions to derive the relation between $\rm M_{Z}/M_{\star}$ and the gas fraction. 
We refer the interested reader to Appendix \ref{app_B} for a detailed discussion of the assumptions, strengths and weaknesses of the different bathtub models and a justification of the choice made in this work.

We further show in appendix \ref{app_A} that the relation between $\rm M_{Z}/M_{\star}$ and $\rm f_{gas}$ is {\it independent} of both the inflow rate and of star formation efficiency and depends only on the outflow loading factor. This is a very important point, since the relation between these two quantities is therefore a clear, non-degenerate tracer of the net effect of outflows (unlike the effective yield). The red lines in Fig.~\ref{model_Mztot} show $\rm M_{Z}/M_{\star}$ as a function of $f_{gas}$, for different values of the outflow loading factor. Coloured symbols in Fig.~\ref{model_Mztot} show the $\rm M_{Z}/M_{\star}$ and $f_{gas}$ observed in different radial annuli in NGC 628, where the colour coding represents the radial distance from the galaxy centre. Filled symbols are for the M08 calibration, while hollow symbols are for the PP04 calibration. We note that this figure is just an alternative representation of the metal deficit discussed in the previous section (Fig. \ref{met_budget}). Here we have replaced the radial distance with $\rm f_{gas}$, but the two quantities are related as discussed in Sec. \ref{grad2}, since the gas fraction increases monotonically with radius. The metal deficit can be visualised in Fig. \ref{met_budget} as the distance between the observational data points and the $\lambda = 0$ (solid red) line, which represents the $\rm M_Z = y \ M_{\star}$ case.

By using the M08 calibration the observed points are well reproduced with an outflow loading factor $\rm \lambda \approx 1$,
with a scatter in the range $\rm 0.5< \lambda <2$, except for the `anomalous' point at 3~kpc, which requires a lower loading factor
approaching zero. The PP04 calibration requires slightly higher loading factors (except again for the anomalous point at 3~kpc).

We emphasise that, since we use the {\it total} mass of metals relative to the mass of stars, our analysis provides an estimate of the {\it average} outflow loading factor during the galaxy lifetime, and not the current outflow rate in the galaxy. Indeed, within the picture of hierarchical formation of structure, it is possible that this average loading factor is mainly driven by the high efficiency of metal loss at higher redshift, when the physical condition in the galaxy and its sub-components would have been significantly different.


\subsection{Modelling the gas metallicity} \label{inflow}

As shown in Appendix \ref{app_A}, the chemical evolution framework adopted in the previous section also leads to a direct relation between gas metallicity ($\rm Z_{g}$) and  gas fraction ($\rm f_{gas}$) that depends only on the outflow loading factor and not on the inflow rate and star formation efficiency. The relation between $\rm Z_{g}$ and $\rm f_{gas}$ has been studied by several authors, including recently \cite{Ascasibar2014} and \cite{Kudritzki2015}, who extend their analysis to resolved scales making use of data from NGC 628. 

However, unlike the metals locked in stars, the gas phase metallicity is more subject to recent gas flow events (possible stochastic variations away from the ideal gas regulatory model), hence making it a less reliable tool for inferring the time averaged properties of the system.

\begin{figure} 
\centering
\includegraphics[width=0.48\textwidth, trim=80 170 120 220, clip]{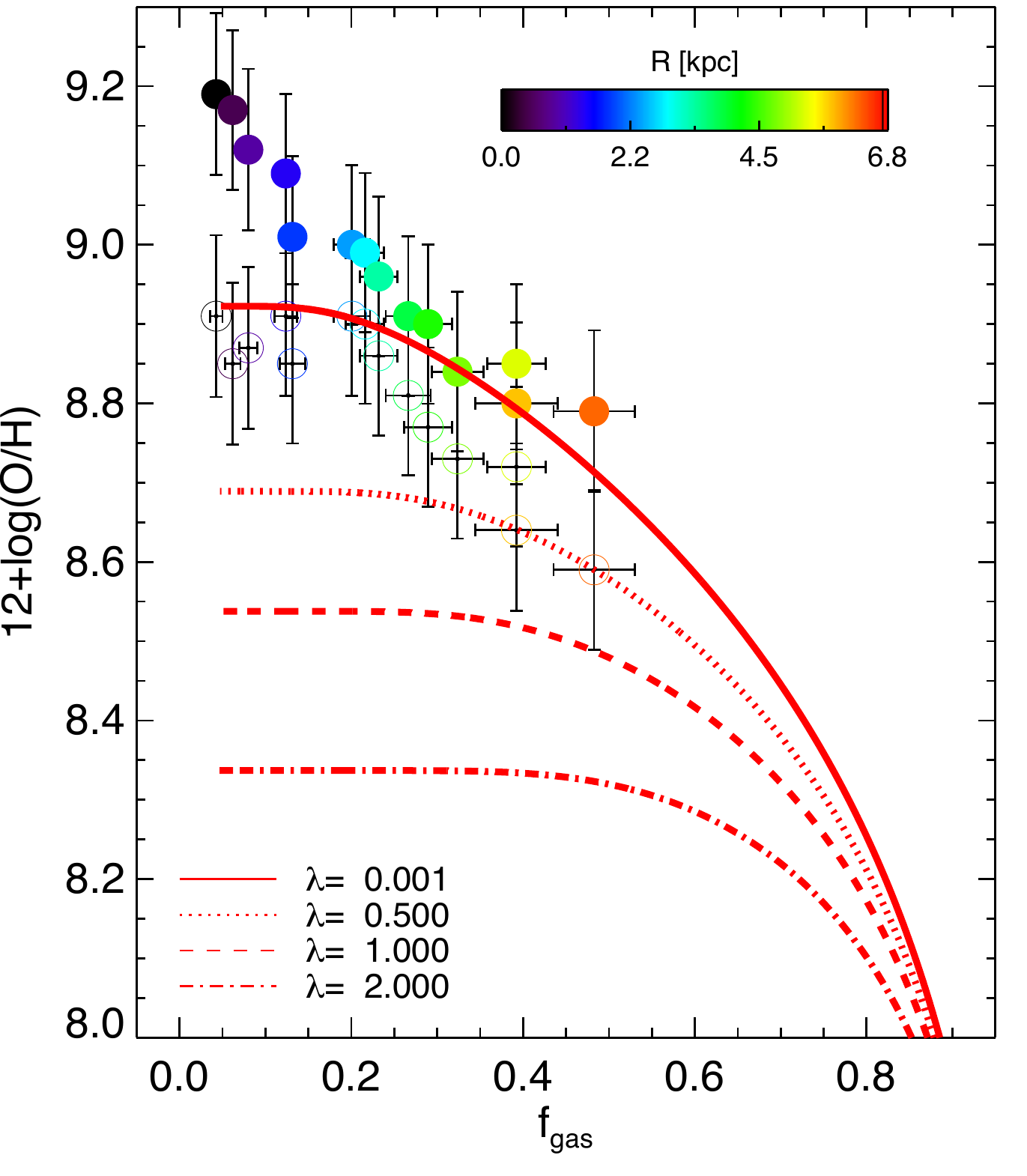}
\caption{Gaseous metallicity $\rm 12+log(O/H)$ versus gas fraction. Symbols show the values observed in radial annuli in NGC 628, where the colour-coding gives the galactocentric radial distance, as indicated in the colour bar. Solid symbols correspond to the mass of metals in which the ISM metallicity is calculated by using the M08 calibration, while open symbols use the PP04 calibration. The red lines correspond to the prediction of simple chemical evolution models with different values for the average outflow loading factor ($\lambda$). Note that the model relations are independent of the inflow rate $\Phi$ and of the star formation efficiency $\varepsilon$.}
\label{model_Zg}
\end{figure}

Keeping these caveats in mind, the model relations between $\rm Z_{gas}$ and $\rm f_{gas}$ are shown in Fig.~\ref{model_Zg}, with the same outflow loading factors as in Fig.~\ref{model_Mztot}. The data points represent the observed values in NGC 628 and are colour-coded by radial distance. The two different metallicity calibrations are identified, as in Fig. \ref{model_Mztot}, by solid \citep{Maiolino2008} and hollow \citep{Pettini2004} circles. The comparison of the models with the data points suggests outflow loading factors consistent with zero, or generally lower than those derived from the $\rm M_Z/M_{\star}$ vs $\rm f_{gas}$ relation (Fig.~\ref{model_Mztot}).

We note that the mismatch applies (with about the same magnitude) both to the central regions, where the total mass of metals is dominated by the stellar component, and in the outer regions, where the total mass of metals is dominated by the gaseous component, therefore excluding that the mismatch is simply associated with a different metallicity scales or calibration issues associated with the calculation of one of the two quantities.

As discussed above, a possible solution can be attained assuming that the gas metallicity might not be representative of the average evolutionary processes during the life of the galaxy since it can be subject to recent galactic `weather'. In particular, a process that may strongly affect the gas metallicity, without significantly affecting the total content of metals, could be recent inflow of enriched gas from the halo. We discuss this option further in Sec. \ref{dis}.

Finally we note that \cite{Kudritzki2015} have inferred a similarly low loading factor ($\rm \lambda$ = 0.2-0.3) for NGC 628, by fitting the gas phase metallicity gradient and the gas fraction. Since they use $\rm T_e$-based metallicities, their results should be compared with our analysis using the PP04 calibration,with which they are in qualitative agreement. However, \cite{Kudritzki2015} make use of a chemical evolution model where the inflow rate is proportional to the SFR. As discussed in Appendix \ref{app_B}, this assumption introduces a dependance of the metallicity-gas fraction relation on the inflow rate, which is absent in our model.

\section{Discussion}
\label{dis}

In the previous section we have demonstrated that a) overall $\sim 50\%$ of the oxygen produced by the NGC
628 out to R = 7.0 kpc is unaccounted for and b) average outflow loading factors of order unity can explain
the observed metal deficit, although there is a tension with the observed gas metallicity. While we cannot entirely exclude that possibility that the tension is due to the unavoidable systematic uncertainties in the determination of gas phase and stellar metallicity and the choice of nucleosynthetic yield/IMF, in this section we discuss possible ways to interpret the discrepancy in light of different physical mechanisms.

\subsection{Gas flows and disc-halo interaction}
It is important to remember that the chemical evolution framework used in this work is an intentionally simple prescription. In particular, while the model allows metal rich outflows and pristine gas inflows, and hence some exchange of gas and metals between the different radial bins, it does not treat the case of incoming metal-rich gas or changes in the stellar mass component due to stellar migration.

A number of physical processes, for example, might be responsible for funnelling metal-rich gas into the
central galactic regions, including viscous flows generated by gravitational instability or cloud - cloud
collisions \citep{Lacey1985, Thon1998, Ferguson2001} or the presence of a bar or other non-axisymmetric
perturbation \citep{Minchev2010, DiMatteo2013}. The breaks seen in the stellar metallicity, and possibly
the central flattening of the gas phase metallicity profile when using the PP04 calibration, combined with
the presence of a star formation ring just inside the break radius, may indeed support the hypothesis of an oval perturbation playing a role in the evolution of the disc in NGC 628.

More importantly, enriched gas expelled by previous generations of supernovae may fall back onto the disc, generating a galactic fountain. In the literature, however, there is no agreement on the details of the fountain mechanism \citep{Melioli2008, Melioli2009, Spitoni2010}, the radial distance covered by a galactic fountain, the time needed for the gas to rain back onto the disc and the overall impact of this mechanism on the chemical abundances of the disc itself.
In hydrodynamical simulations, if a hot halo is present, metal-enriched gas can fall back towards the galactic centre \citep{Melioli2009}. 

The presence of large amounts of metals in the halo of disc (and elliptical) galaxies has been inferred by
recent HST-COS observations of absorption systems along the line of sight of background quasars
\citep[e.g.][]{Tumlinson2011,Werk2014}. However, these observations have probed the halo of galaxies at large radii ($\sim$100~kpc). In the inner region of the halo the gas metallicity is not known observationally, but simulations expect that the metallicity of the diffuse gas should reach values higher than solar \citep{Marinacci2014}.

\begin{figure*} 
\centering
\includegraphics[width=0.9\textwidth, trim=30 90 80 100, clip]{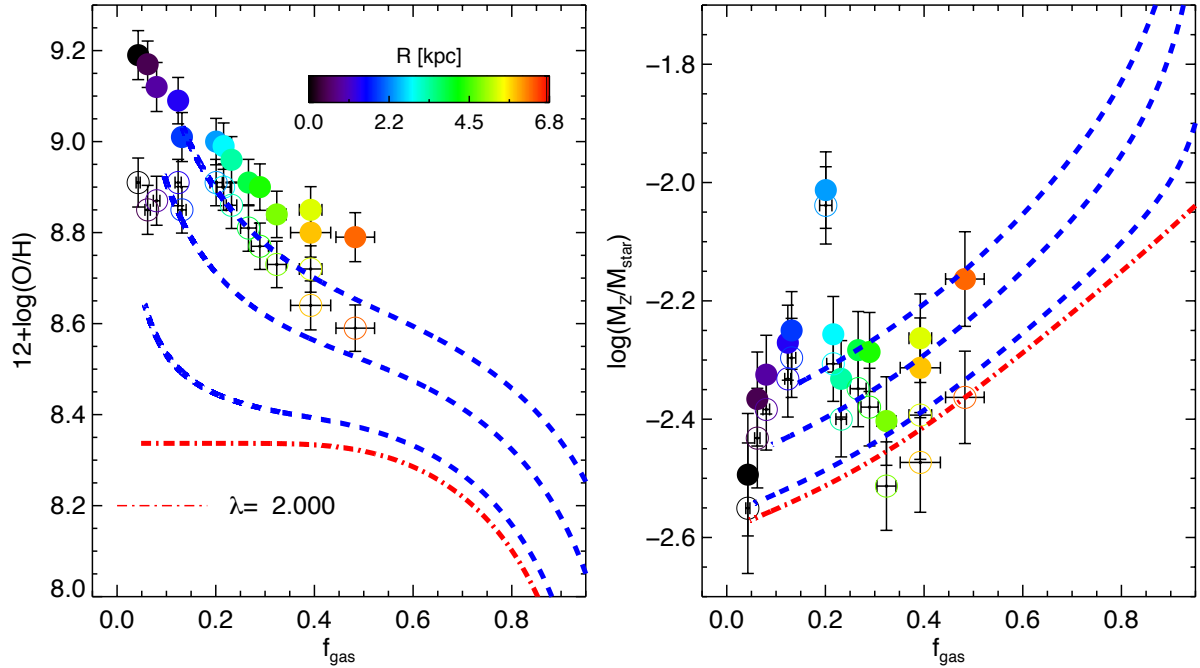}

\caption{Same as Fig. \protect\ref{model_Mztot} and \protect\ref{model_Zg}, but including a late time inflow of gas rich material with metallicity $\rm 3 \ Z_{\odot}$. The red line is the same as in Fig. \protect\ref{model_Mztot} and \protect\ref{model_Zg} and represents the model with $\rm \lambda = 2$, while the blue lines represent an increasing mass of infalling metal-rich gas. Going upwards in both figures the blue lines represent a infalling metal-rich gas mass that is a fraction of 1$\%$, 5$\%$ and 10$\%$ of the mass of baryons already present in the disc at a given location.}
\label{model_inflow}
\end{figure*}

We have attempted to model an episode of metal-enriched accretion at late times by modifying the analytical
description of the metallicity evolution discussed above with the introduction of an accreted enriched gas
mass, proportional to the mass of baryons already present in the galaxy, and with a given metallicity $\rm
Z_{gas-accr}$. The blue lines in Fig.~\ref{model_inflow} shows the effect of such a late enriched accretion
by assuming that the accreted gas is 1, 5 and 10\% of the mass in baryons already present in the radial bin
and that the metallicity of such enriched inflow is $\rm Z_{gas-accr} = 3~Z_{\odot}$ \citep[which is
consistent with the enrichment of the hot gas phase expected by some models,][]{Marinacci2014}.
Clearly such a model can now broadly reproduce the trend of the gas metallicities in NGC 628 at several radii.
At the same time, Fig.~\ref{model_inflow} illustrates that such a late enriched accretion has little effect
on the total content of metals, hence having little effect on the global, average outflow rate that we have
inferred from the metal deficit.

In these models we have assumed an outflow loading factor $\lambda=2.0$, i.e. higher than inferred in the previous section, to better illustrate that late enriched accretion can have a strong effect on the final gas metallicity but little effect on the total content of metals.
We note that by starting with a lower loading factor (e.g. $\lambda=1.0$) the enriched
inflow can reproduce the data with a smaller amount of accreted gas ($\sim 3-5$\%).

Both gas flows in the disc and galactic fountains might therefore help resolve the tension between the metal deficit and the outflow loading factor implied from the gas phase metallicity for the central regions of NGC 628 by increasing the metallicity of the central regions beyond what is expected in the absence of these metal rich inflows. We note that there is a fundamental degeneracy in this type of model between the outflow loading factor and the mass (and metallicity) of the enriched inflow when using the gas phase metallicity alone. However better constrains can be derived if one models, as we do here, both the stellar and the gas phase metallicity at the same time, because the gas phase metallicity is largely independent of gas flows at late times.

Stellar migration \citep{Roskar2008a, Yoachim2012, Spitoni2015} might also play a role in alleviating the observed tension. If stars diffuse out of the radial bin in which they are born, we would be making incorrect inferences for the gas fraction and oxygen mass in the bin. We can model this effect naively by considering that the observed stellar mass in each bin should be corrected for the net effect of migration. We experimented with this prescription and found that changes in the stellar mass up to 0.3 dex only generate small changes in $\rm M_Z/M_\star$ for all bins within R = 5 kpc, since the metal budget is dominated by the stellar component and therefore $\rm M_Z/M_\star$ is only weakly dependent on stellar mass. At larger radial distances the oxygen mass budget is dominated by the ISM and the effect of changing the stellar mass content is larger. If a large fraction of the stellar mass currently observed at large radii is the result of stellar migration from the centre, we would require a lower outflow loading factor in the outer disc, going towards a reduction of the tension between the metal deficit and the observed gas metallicity.

\subsection{Towards larger galactocentric distances}

\label{large_dist}

The behaviour of the metal budget at even larger radii than the ones probed by our investigation constitutes a particularly interesting avenue for future work. Since H\textsc{i} discs extend substantially more than the stellar component, low levels of chemical enrichment in the outer disc might mean that a large fraction of metals have escaped our budget. Unfortunately there is no consensus in the literature on the nature of the metallicity gradient for $\rm R > 2 \ R_e$. While \cite{Bresolin2012} and \cite{Sanchez2014} argue for a flattening of the gradient beyond $\rm R= 2 \ R_e$, \cite{Moran2012} observed a sharp drop in the metallicity gradient for a subset of H\textsc{i}-rich galaxies. 

Direct measurements of the chemical abundances at large galactocentric radii have been recently obtained for NGC 628 by \cite{Berg2015}. Their observations contain 12 H\textsc{ii} regions at radii larger than those sampled by the IFS data presented in this work. With the very scarce and scattered available data, there is no evidence for either a flattening or a break in the metallicity gradient of NGC 628 out to 10 kpc. This observations are interesting in the context of the metal budget because they confirm the trend that regions at large galactocentric distances have retained a larger fraction of their metals compared to the inner regions of galaxies. In particular, relying on the metallicity gradient reported in \cite{Berg2015} we conclude that the radial bin around 10 kpc has retained $\sim$80-100\% of the metals produced in situ.

Intriguingly, previous work has also presented evidence that H\textsc{ii} regions at extreme galactocentric distances ($\rm R > R_{24} \approx 3 \ R_e$) are more metal rich than expected from in situ star formation \citep{Bresolin2009, Bresolin2012}. These observations might be consistent with an extrapolation of the trend discussed above and in general with the idea that gas rich, relatively un-evolved systems have a smaller metal deficit, or might even have acquired metals through enriched inflows from the central regions (through a galactic fountain). In the context of inside-out disc growth, these outer regions are still in the process of assembly and thus represent an ideal laboratory to study disc formation as it happens.

\subsection{The effect of the IMF and the nucleosynthetic yield}

In this work we have used the \cite{Kroupa1993} IMF, since this is the IMF which fits best the chemical evolution of galaxy discs \citep{Romano2005, Romano2010}. 
A different IMF would have an impact on our calculations by implying a different population-averaged net yield, since different elements are produced by stars of different masses.
In this section we briefly discuss the implications of adopting different IMFs.

A Salpeter IMF would give a significantly higher metal deficit, implying a loss of about 75\% of the metals produced. The inferred average loading factor would be about $\lambda \sim 2-4$. In the case of a \cite{Chabrier2003} IMF
the metal deficit is even more severe: more than 85\% of the metals should be lost and in the implied average
loading factor should be in the range $\lambda > 5$. The \cite{Kroupa2001} IMF gives results very similar to \cite{Chabrier2003}. We note that in going from a Kroupa to a Salpeter IMF we correct the stellar mass profile by a factor of 0.26 dex.

Such large loading factors are extreme, especially considering that these are loading factors averaged during the lifetime of the galaxy, and have never been seen in galactic winds studies locally or even in high-z galaxy discs \citep[e.g][]{Steidel2010,Genzel2011}, which instead infer outflow loading factors more in line with the results obtained by using the \cite{Kroupa1993} IMF, further confirming that this IMF is the most appropriate for modelling galaxy discs.

However, it has been pointed out that the chemical properties of spheroids (including bulges) may be more properly described by a \cite{Chabrier2003} IMF. If we apply the latter IMF only to the central region, we get a large metal deficit ($\sim 90$\%) for the bulge. This would be in line with a similar estimates for the Milky Way, according to which the bulge of our Galaxy has lost about 80\% of its metals \citep{Greggio2011}. The \cite{Chabrier2003} applied to the central region would also imply a central outflow loading factor of about $\lambda \sim 5$. Since we are only measuring the time-averaged outflow loading factor, it is not implausible to find that the central regions of galaxies might have experienced stronger outflows, despite the fact that at present time they sit at the bottom of the galaxy's potential well.
Indeed outflow rates with such large loading factor are obseerved in the central region of local and high-z galaxies \citep{Feruglio2010,Rupke2011,Maiolino2012, Cicone2014, Cicone2015}, and may be explained with the additional boost to the outflow rate associated with quasar-driven winds or with extreme nuclear starbursts in the past, when these feedback processes were likely more violent than in the present Universe.

\subsection{The oxygen deficit in context}
Previous work has estimated the oxygen (or metal) budget in star forming disc galaxies by statistical arguments, making use
of scaling relations (between mass and SFR, between gas fraction and stellar mass, and between mass and
metallicity) demonstrated to hold in large samples of galaxies observed spectroscopically
\citep{Bouche2007, Zahid2012, Peeples2014}. These works predict an oxygen deficit ranging from $35\% $ to
$90 \%$, with large statistical uncertainties, related both to the intrinsic scatter in the scaling
relations used as input for the models and to the difficulty to robustly correct for aperture effects.
Nonetheless it is rewarding to note that our estimate for the metal deficit in NGC 628 falls inside the
range of estimates from previous works. 

Due to the difficulty of studying the stellar metallicity and thus deriving the total metal content in
galaxies, the relation between gas phase metallicity and gas fraction has been widely used as an observationally more accessible tool to investigate the impact of outflows. Recent work \citep{Lilly2013, Dayal2013, Ascasibar2014, Lu2015} generally agrees on loading factors of order unity for galaxies of stellar mass $log(M_\star/M_\odot) \sim 10$.

Extending this study to a larger sample of galaxies with resolved gas masses and metallicity information
would provide the logical way forward to studying the effect of outflows across the whole galaxy population
and uncover trends with galaxy parameters (stellar mass, SFR etc). In particular, low mass galaxies ($\rm log(M_\star/M_\odot) < 9$) are predicted to exhibit higher outflow loading factors in order to match their low observed baryon fractions \citep{McGaugh2010}, their metallicity \citep{Lu2015} and to reconcile the faint-end slopes of the dark matter and galaxy mass functions \citep{Lilly2013}. Some observational evidence for higher outflow rates in dwarf starbursts can be found in the literature \citep[e.g][]{Martin2002}, but more evidence in this mass regime is urgently needed.

While such a program is observationally costly, the synergy between large IFS surveys (CALIFA, SAMI and MaNGA) and radio and sub-mm interferometers (ALMA, NOEMA, JVLA) will greatly increase the availability of matched-resolution datasets to use for this type of study in the near future.

\subsection{A link between outflows and baryon fraction}

The metal deficit is tightly related to the well-known `missing baryon problem' \citep[e.g.][]{Fukugita1998, McGaugh2010, Shull2012}. Within our `baryon cycling' framework we cannot account for the inventory of cosmic baryons, a large fraction of which are believed to exist in the IGM and circumgalactic medium. However, within the gas regulatory model, outflows modulate the final baryonic mass of the system, while the gas inflow rate can be assumed to be directly proportional to the dark matter halo mass growth rate. \cite{Peng2014} provide a useful parametrisation to relate the halo mass ($\rm M_h$) and the gas inflow rate ($\Phi$)

\begin{equation}
\Phi = f_{b}~f_{gal}~\frac{dM_h}{dt}, \label{dark1}
\end{equation}

where $\rm f_b = 0.155$ \citep{Planck2013} is the cosmic baryon fraction and $\rm f_{gal}$ is the fraction of accreting baryons that penetrates into the inner galaxy disc, which could be a complex function of halo mass and redshift. Relying on results from the hydrodynamical simulations in \cite{Dekel2013}, we follow \cite{Peng2014} and assume for simplicity $\rm f_{gal} \sim 0.5$ in this section. 

The evolution of the total baryonic content of the system ($\rm M_b = M_g + M_\star$) is the dictated by Eq. \ref{eqa1}--\ref{eqa2} so that $\rm dM_b/dt = \Psi - \Phi$. Combining this with Eq. \ref{dark1} we obtain 

\begin{equation}
M_b = f_b~f_{gal}~M_h - \frac{\lambda}{1-R}~M_\star,
\end{equation}

Rearranging we finally obtain a relation between gas fraction ($\rm f_{gas}$), outflow loading factor ($\rm \lambda$) and the detected baryon fraction ($f_d \equiv \frac{M_b}{f_b~M_h} $, as defined in \citealt{McGaugh2010})

\begin{equation}
f_d \equiv \frac{M_b}{f_b~M_h} = f_{gal} \left( 1 + \frac{\lambda}{(1-R)(1+f_{gas})} \right) ^{-1}.
\end{equation}

Using representative values for NGC 628, $\rm f_{gas} = 0.2 $ and $\rm \lambda =1$ we obtain $f_d \sim 0.23$. This very rough order-of-magnitude calculation leads to a detected baryon fraction that is surprisingly close to the detected baryons fractions measured via the Tully-Fisher relation in \cite{McGaugh2010}, who quote $f_d \sim 0.16-0.19$ for disc galaxies of stellar mass $\rm log(M_\star/M_\odot) = 9.9 -10.1$, comparable to NGC 628.

\section{Conclusions}
\label{concl}

In this work we have presented a detailed study of the metal content and distribution in the nearby disc galaxy NGC 628. We have mapped the gas phase metallicity by exploiting the largest integral field spectroscopy mosaic available to date, thus enabling us to extend our study out to a galactocentric radius of 7~kpc ($\rm \sim 3 \ R_e$). We have combined the gas metallicity map with the stellar metallicity radial profile presented in previous work \citep{Sanchez-Blazquez2014}, the stellar mass surface density distribution $\Sigma_\star$ (inferred from extensive multi-band photometry) and the observed gas surface density ($\rm \Sigma_{HI}, \Sigma_{H2}$) inferred from CO and H\textsc{i} maps. 

By comparing the mass of metals observed in the gaseous and stellar components with the mass of metals produced by stars in the same region (inferred assuming the latest determination of the oxygen nuncleosynthetic yield, \citealt{Vincenzo2015a}) we have obtained a detailed, spatially resolved metal budget as function of galactocentric radius. The main results from this analysis are the following:

\begin{enumerate}

\item On average about 50\% of the metals produced by the stars in each galactic region have been lost.
The fraction of metals lost is higher ($\sim$70\%) in the central region, which is dominated by the bulge. There is also tentative indication that the fraction of metals lost decreases at large galactocentric radii. Cumulatively, out to a radius of 7~kpc, 45\%--50\% of the {\it total} amount of metals produced have been lost by the galaxy. Evidence from recent measurements of outlying H\textsc{ii} regions from \cite{Berg2015} (out to 10 kpc) confirms the trend that regions in the outer disc have retained a higher fraction of the metals they have produced, with regions at around 10 kpc retaining $\sim$ 80-100 \% of the metals produced there.

\item We have used simple (`bathtub' and `gas regulatory') analytical models involving gas outflow and inflow to model the data. In this framework the relation between metals-to-stellar mass ratio ($\rm M_Z/M_{star}$) and gas fraction ($\rm f_{gas}$) is independent of the inflow rate and of the efficiency of star formation, and depends only on the outflow loading factor in models with constant inflow rate. We find that an outflow loading factor $\lambda \approx 1$ can explain the data at most radii, though with large scatter ($0.5< \lambda < 2$). We emphasise that, since this inference is based on the total metal budget, we derive a time-averaged outflow rate, representative of the strength of outflows over the lifetime of the galaxy. This information is thus complementary to that obtained in studies based on the observed gas kinematics, which are sensitive only to the ongoing/recent outflow activity.

\item A larger loading factor may be derived if we make use of the Chabrier IMF. This IMF might be relevant for the central spheroid, thus implying a higher outflow rate during the early formation of the bulge.

\item The gas phase metallicity is more sensitive to recent evolutionary processes because it is associated with a component of the galaxy (the ISM) that is potentially subject to rapid evolution. We find that the observed gas metallicity cannot be reproduced by the same parameters inferred from the metal budget, independently from the choice of metallicity calibrator used. In particular, the observed gas phase metallicities would require little or no outflow. We show that the apparent tension could be explained with infall of enriched gas from the halo at late times. Indeed, such a process would strongly affect the observed gas metallicity, but not the total mass of metals.

\item Finally, we warn against the use of analytical models in which the inflow rate is assumed to be proportional to the star formation rate. This assumption, adopted in previous work, is not physically motivated and introduces an unnecessary degeneracy between inflow and outflow rates when modelling chemical abundances. Analytical solutions are presented in this work for models assuming a constant, or slowly varying, accretion rate, which is a much better representation of gas inflows in the cosmological framework. However, further work is needed to reliably test the validity of the underlying assumptions of the simple regulatory model presented in this work for describing regions within galaxies.

\end{enumerate}

\section*{Acknowledgment}
FB acknowledges support from the United Kingdom Science and Technology Facilities Council.
This work makes use of THINGS (`The Nearby Galaxy Survey', \citealt{Walter2008}), HERACLES (the `HERA CO line Extragalactic Survey', \citealt{Leroy2009}) and PINGS (the `PPAK IFS Nearby Galaxy Survey', \citealt{Rosales-Ortega2010}). We acknowledge the contribution from the referee in improving the content and the clarity of the paper. We thank Fiorenzo Vincenzo and Francesca Matteucci for their invaluable help in interpreting the vast literature on chemical abundance modelling and nucleosynthetic yields. We also thank Ying-jie Peng for support and discussion on development of his chemical evolution models. We wish to thank Fabian Rosales-Ortega for kindly sharing the PINGS data on NGC 628 and for his encouragement and feedback on the early stages of this work. We thank Matt Auger for useful discussions and healthy skepticism and the participants to the MPIA summer workshop `A 3D view on galaxies evolution: from statistics to physics' for insightful comments.

\bibliography{ngc_628_2.bib}
\bibliographystyle{mn2e}

\appendix

\section{Analytical solution for constant inflow models}
\label{app_A}

In this section we summarise the chemical evolution formalism of \cite{Peng2014} and present further exact solutions for the metallicity and total metal content.

\cite{Peng2014} models offer solutions for equations Eq. \ref{eqa1}--\ref{eqa4} under the following assumptions:
\begin{enumerate}
\item{Constant inflow rate $\Phi$.}
\item{A linear relation star formation law $\rm SFR = \varepsilon ~ M_{gas} $, with constant $\varepsilon$.}
\item{Outflow rate proportional to the SFR through a constant loading factor $\lambda$: $\rm \Psi = \lambda ~SFR$.}
\end{enumerate}

We further assume that the net nucleosynthetic yield p and return factor R are constant with time and the the inflow is of primordial gas (with zero metallicity). Under these assumptions analytical solutions can be obtained with an explicit time dependance. As discussed in \cite{Peng2014} the equilibrium timescale 

\begin{equation}
\rm \tau \equiv \frac{1}{\varepsilon~(1-R+\lambda)}
\label{teq}
\end{equation}

is the natural timescale driving the chemical evolution of the system. For $\rm t >> \tau$ physical properties of galaxies tend to their `equilibrium' values. As discussed in \cite{Peng2014}, low mass dwarf galaxies and chemically un-evolved systems will not satisfy the equilibrium condition and need to be studied in the $\rm t < \tau$ regime.

In Table \ref{table_sol} we summarise the analytical solutions for a number of galaxy properties. We note that, apart from the assumptions stated above, the solutions are mathematically exact\footnote{The solution to the gas phase metallicity ($\rm Z_g$) presented here differs from the one in \cite{Peng2014}, who solve for the metallicity evolution making further approximations to simplify the algebra.}. We also remark that, as expected, the only timescale responsible for the chemical evolution of the system is the equilibrium timescale defined above (Eq. \ref{teq}).

The limits of the solutions for the equilibrium case ($\rm t >> \tau$) have a simple physical interpretation. Star formation exactly balances out the inflow and outflow rates, the gas mass (and hence the SFR) stays constant and the stellar mass grows linearly with time. In this regime the gas phase metallicity plateaus at a constant value

\begin{equation}
Z_g (eqm) = p~\varepsilon~\tau = \frac{p}{1-R+\lambda},
\end{equation}

which depends only on the net yield, the return fraction and the outflow loading factor. 
Note also that the equilibrium value of the $\rm M_Z/M_\star$ ratio is $\frac{p}{1-R} (1-\varepsilon~\lambda~\tau) $, lower than the expected value $\rm M_Z/M_\star = \frac{p}{1-R}$ in absence of outflows (i.e. for $\rm \lambda = 0$). The gas fraction $\rm f_{gas}$ can be considered as a function of $\rm t, \tau$ and $\varepsilon$, and is the best property to quantify the degree of chemical evolution of the system: systems with low gas fraction have not reached equilibrium and are chemically un-evolved, while systems with low gas fraction are in equilibrium and chemically evolved.

The analytical solutions are also an important tool to study the parameter dependences of different physical properties. In particular we can easily show that the relation between $\rm M_Z/M_\star$  and $\rm f_{gas}$ depends only on the value of the outflow loading factor, and does not depend on the assumed inflow rate and star formation efficiency. In other words the position of a point in the  $\rm f_{gas}$ vs $\rm M_Z/M_\star$ plane depends only on $\lambda$. 
To prove this it is sufficient to observe that both $\rm f_{gas}$ and $\rm M_Z/M_\star$ are only a function of $\rm \lambda$ and $\rm t/\tau$. Therefore, eliminating $\rm t/\tau$, it would be possible to write a relation between $\rm f_{gas}$ and $\rm M_Z/M_\star$ which depends only on $\rm \lambda$. Unfortunately the complex nature of the solutions does not allow such relation to be written in closed form. Physically this statement can be interpreted as stating that the metal deficit (i.e. $\rm M_Z/M_\star$) depends only on how evolved the system is (i.e. its $\rm f_{gas}$) and the strength of its outflows (i.e. $\lambda$). As shown in App. \ref{app_B}, this simple result is \textit{not} valid anymore if we assume that the inflow rate is proportional to the SFR. 

Finally we note that exactly the same argument can be applied to the relation between $\rm f_{gas}$ and $\rm Z_g$, which can therefore be shown to depend only on the outflow loading factor and not on the inflow rate and the star formation efficiency.

\begin{table*}
\caption{Exact analytical time-dependent solution of the \protect\cite{Peng2014} models with constant inflow rate for different galaxy properties. The equilibrium timescale $\tau$ is defined in Eq. \ref{teq}.}

\begin{tabular}{ l c c c}
\hline
Galaxy property 	& 	Exact solution & 	$\rm t << \tau$ & 	$\rm t >> \tau $ \\
\hline

$M_{g} $ & $ \Phi~\tau~(1-e^{-t/\tau}) $	&	$ \Phi~t $	& $ \Phi~\tau $ \\
\\

$SFR = \varepsilon~M_{g} $ & 		$ \varepsilon ~\Phi~\tau~(1-e^{-t/\tau}) $			
&		$ \varepsilon~\Phi~t $		& $  \varepsilon~\Phi~\tau$ \\
\\

$M_{\star} $ & 		$ \Phi~\tau^2~\varepsilon~(1-R)~(t/\tau - (1- e^{-t/\tau}) ) $			
&	$ \Phi~\varepsilon~(1-R) ~t^2 /2 $			& $ \Phi~\varepsilon~(1-R)~\tau ~t$ \\
\\

$f_{gas} \equiv \frac{M_{g} }{M_{g} +M_{\star}} $ & 		$ \left[ 1 + \varepsilon~ \tau~(1-R) ~ \left( \frac{t}{\tau} ~ \frac{1}{1- e^{-t/\tau}}  -1 \right) \right]^{-1} $  	& 	$(1+\varepsilon~ (1-R)~t/2)^{-1}$  & 	$(1+\varepsilon~ (1-R)~t)^{-1}$  \\

\\

$Z_{g} \equiv M_{Zg}/M_{g} 	$	&  	$p~\varepsilon~\tau \left( 1 - \frac{t/\tau ~ e^{-t/\tau} } {1- e^{-t/\tau}} \right)$ &
$ p~\varepsilon~t/2$ & $p~\varepsilon~\tau$ \\
\\

$M_Z	$	& 	$p~\varepsilon~\Phi~\tau^2 \left[ (1-\epsilon~\lambda~t) (t/\tau - ( 1- e^{-t/\tau})) - \varepsilon~\lambda~t ( (t/\tau + 1) e^{-t/\tau} -1 ) \right] $ & $p~\varepsilon~\Phi~t^2/2$ & $ p~\varepsilon~\Phi~(1-\varepsilon~\lambda~t)~\tau~t$ \\
\\

$M_Z/M_\star$ & $\frac{p}{1-R} \left[ 1- \varepsilon~\lambda~\tau \left( 1+ \frac{ ( (1+t/\tau  ) ~e^{-t/\tau} -1 )} {t/\tau + e^{-t/\tau}  -1} \right) \right] $  & $\frac{p}{1-R}$ & $\frac{p}{1-R} (1-\varepsilon~\lambda~\tau) $ \\
\hline

\end{tabular}


\label{table_sol}
\end{table*}

\section{Equilibrium, timescales and inflow rates in different `bathtub' models}
\label{app_B}

In this section we briefly recap the main features and differences between the growing set of chemical evolution `bathtub' models. We will focus on a comparison of the models presented in \cite{Dave2012}, \cite{Lilly2013}, \cite{Dayal2013} and \cite{Peng2014}, the latter being our reference model. The casual reader is encouraged to read the summary at the end of this section.

A `bathtub', or `gas regulatory' model, is defined here to be a system assumed to obey the equations  \ref{eqa1}--\ref{eqa4}. An `ideal' bathtub model is one in which we assume 

\begin{enumerate}
\item{A linear relation star formation law $\rm SFR = \varepsilon ~ M_{gas} $, with constant $\varepsilon$.}
\item{Outflow rate proportional to the SFR through a constant loading factor $\lambda$: $\rm \Psi = \lambda ~SFR$.}
\end{enumerate}

The simplest bathtub model, like the one presented in \cite{Dave2012}, is an `equilibrium' model. An equilibrium model is defined by the assumption that the gas mass in the system does not change with time ($\rm dM_{gas} /dt = 0$). This assumption is justified by \cite{Dave2012} on grounds of its relevance to cosmological hydrodynamical simulations. 
In equilibrium Eq. \ref{eqa1} reduces to 

\begin{equation}
\Phi  = (1-R+\lambda) \ SFR  \quad \text{ (in eqm)}. \label{eqaa1}
\end{equation}

Note that even in an equilibrium model the SFR will change in time, however equilibrium requires that it does so at the same rate as the inflow rate. The same conclusion is evident in the equilibrium solution for the SFR presented in Table \ref{table_sol} above.

 \cite{Dave2012} also argue that, when departures from equilibrium are allowed, the bathtub model naturally generates a secondary dependance of the mass-metallicity relation on SFR, in the sense that more gas rich systems will have higher SFR and lower metallicities. This feature of the model is generally considered a success \citep{Lilly2013}, since it is capable of reproducing the observed fundamental metallicity relation \citep{Mannucci2010, Andrews2013}. Moreover this feature is mostly independent of the detailed time-evolution of the cosmological inflow rate, since it is just a general characteristic of the non-equilibrium bathtub models.
 
Subsequent work in \cite{Lilly2013} has shown that equilibrium is generally a good assumption for gas-poor, chemically evolved, massive low-redshift galaxies, while it is not a good assumption for the gas-rich dwarfs or galaxies at high redshift (see in particular Sec. 2.3 and 2.4 of \citealt{Lilly2013} for a detailed discussion of the relevant timescales and Sec. 5 of \citealt{Peng2014} for an estimate of the equilibrium timescale for gas rich systems). It is therefore clear that in applying the bathtub model to the chemically un-evolved, gas rich outer disc of NGC 628 a model that takes into account departure from equilibrium is needed.
 
The simplicity of the framework offered by the bathtub model has encouraged several authors to seek analytical solutions for its fundamental equations. Unfortunately, if we do not assume the system is in equilibrium, the relation between metallicity and gas fraction (which can be interpreted as natural `clock' in the context of chemical evolution) will depend on the assumed time evolution of the inflow rate. 

A notable special case is represented by the popular solutions for the case in which in the inflow rate is proportional to the SFR \citep{Matteucci1983, Dayal2013, Kudritzki2015}. The success of these solutions is to be attributed to the fact that they are arguably the simplest example of non-equilibrium models (i.e. $\rm dM_{gas} /dt \neq 0$) which have closed-form analytical solutions. As all non-equilibrium bathtub models, they show the usual behaviour of tending to equilibrium at late times and predict a secondary relation on SFR in the mass-metallicity relation \citep{Dayal2013}. However, we warn the reader against fallaciously reversing the argument, i.e. the existence of the fundamental metallicity relation does not necessarily imply that the inflow rate must be proportional to the SFR!

It is important to stress that, while the other assumptions made in the context of the bathtub model can be justified, albeit approximately, on physical or observational grounds, \textit{there is no physical or observational reason to believe that the inflow rate should follow the same time evolution as the SFR when the system is not in equilibrium}. Confusion often arises in the literature because, in equilibrium, all bathtub models predict the time evolution of the SFR to the follow that of the inflow rate. However, this conclusion is only true when the system is at equilibrium and \textit{not} true in general as the system approaches equilibrium. 

On the other hand, the cosmological inflow rate follows a well-defined trend with halo mass and redshift. Both semi-analytics and cosmological simulations demonstrate that the halo accretion rate can be parametrised as 

\begin{equation}
\frac{dM_{h}}{dt} = \alpha M_{h} (1+z)^\beta, \label{eqaa2}
\end{equation}

where $M_{h}$ is the halo mass and $\rm \alpha \sim 0.030 \ Gyr^{-1}$ and $\rm \beta \sim 2.5$ \citep{Dekel2013}. For an Einstein-de-Sitter universe the Eq. \ref{eqaa2} admits a particularly simple solution leading to

\begin{equation}
M_{f}=M_{i} e^{-a (z_f - z_i)},  \quad  a=\alpha/(H_0 \Omega_m^{1/2}) \sim 0.8
\end{equation}

where the subscripts f and and i refer to final/initial masses and redshifts. Therefore, by substituting back into Eq. \ref{eqaa2}, the halo mass accretion rate at redshift z, $\dot{M_{h}}(z)$ is given by

\begin{equation}
\dot{M_{h}}(z) = 0.03 \ M_i \ e^{-a (z_f - z_i)} (1+z)^{2.5}
\end{equation}

As reported in previous work \citep{Dekel2013, Lilly2013}, \textit{this halo accretion rate does not evolve very much over most of the age of the Universe}. In particular, it changes by about a factor of 2 between z = 5 and 0.3. We neglect here that fact than only a fraction of accreting baryons will reach the galaxy disc and are thus available for star formation, and this fraction might be a complex function of mass, redshift and perhaps other quantities (environment). Under these assumptions, \cite{Peng2014} argue that the inflow timescale is much longer than all other timescales relevant to galaxy evolution (including the equilibrium timescale) and, therefore, it is an excellent approximation to assume the gas inflow rate to be constant in time. By making the assumption that the timescale for changes in the infall rate is much longer than the equilibrium timescale we obtain the models presented in Table \ref{table_sol}, which have been used throughout this paper.

To demonstrate the physical difference between the solutions that assume a constant inflow rate and the solutions which assume the inflow rate to the proportional to the SFR at all times, we present a comparison of their time evolution. The first thing to appreciate is that, if we assume the inflow rate to the proportional to the SFR ($\rm \Phi = \omega~SFR$), Eq. \ref{eqa1} only admits the following solution for the gas mass $\rm M_g$

\begin{equation}
\frac{dM_g}{dt} = (\omega - \lambda - 1+R) \varepsilon M_g \rightarrow M_g=M_0 e^{- (1-R -\omega + \lambda) \varepsilon t},
\end{equation}

where $M_0$ is the initial gas reservoir at t=0. Physical solutions require $\rm 1-R -\omega + \lambda>0$, and hence imply an exponentially declining SFR. The gas equilibrium timescale ($\tau'$) for this model depends on $\omega$ and is given by

\begin{equation}
\tau' = \frac{1}{\varepsilon (1- R + \lambda - \omega)}. \label{eqaa3}
\end{equation}

This timescale tends to the timescale in \ref{teq} for small $\omega$. We therefore refer to the model with SFR proportional to the inflow rate as an $\omega$-model. The equilibrium behaviour of the $\omega$-model is very different from the constant inflow regulator. In the $\omega$-model equilibrium can only be reached for $\rm SFR = \varepsilon~ M_g = 0$. Because of its defining assumption, this model implicitly assumes the inflow rate ($\rm \Phi = \omega ~SFR$) to be exponentially declining (unless we wish to build a more general model with a different star formation law, i.e. $\rm SFR \neq \varepsilon ~M_{g}$, which we will not discuss here). In light of the discussion on the cosmological accretion rate above, this is a much worse parametrisation than the assumption of a constant inflow rate. 
However, an exponentially declining inflow rate is still a useful model in the context of the overall evolution of galaxies, since it can represent the shutdown of star formation due to a declining inflow rate. This model is therefore similar to the `strangulation' models used in \cite{Peng2015} to study the build up of the red sequence.

Although the relation between metallicity and gas mass in the $\omega$-model has been widely used in the literature, it is instructive to present here the solutions for the time evolution of this model, which can be found in Table \ref{table_sol2}. We note that, unlike in the constant inflow model, the metallicity in the $\omega$-model does not come to equilibrium on the gas equilibrium timescale defined in \ref{eqaa3} but on the dilution timescale $\tau_{dil} \equiv (\varepsilon~\omega)^{-1} = M_g/\Phi$. Therefore the final relation between metallicity and gas content will depend on combination of the gas equilibrium and the dilution timescales. In particular, by combining the solutions for $\rm Z_g$ and $M_g$ we obtain the well-known relation

\begin{equation}
Z_g = \frac{p}{\omega} \left( 1 - \left( \frac{M_g}{M_0} \right) ^{\omega/(1-R-\omega+\lambda)} \right),
\end{equation}

and by combing the solution for $M_\star$ and $M_g$ we can eliminate $M_0$ from the above the re-write the metallicity in the $\omega$-model completely in term of observables.

\begin{equation}
Z_g = \frac{p}{\omega} \left( 1 - \left( \frac{M_\star}{M_g} (1+w) \right) ^{-\omega/(1-R-\omega+\lambda)} \right),
\end{equation}

where $w\equiv(\lambda - \omega)/(1-R)$. As can be explicitly seen from this equation, \textit{the relation between $Z_g$ and gas fraction depends on the inflow rate (through $\omega$), causing the observed degeneracy in the determination of $\omega$ and $\lambda$ when using this model}.

\begin{table*}
\caption{Exact analytical time-dependent solution of the ideal regulator with inflow rate proportional to the SFR ($\rm \Phi = \omega ~ SFR$) for different galaxy properties. Note that $M_0$ below refers to the gas mass at t=0 and $\tau'$ is the gas equilibrium timescale defined in Eq. \ref{eqaa3}.}

\begin{tabular}{ l c c c}
\hline
Galaxy property 	& 	Exact solution & 	$\rm t=0$ & 	$\rm t >> \tau'$ \\
\hline

$M_{g} $ & $ M_0~e^{-t/\tau'}$	&	$ M_0 $	&  0  \\
\\

$SFR = \varepsilon~M_{g} $ & 		$ \varepsilon ~M_0~e^{-t/\tau'} $			
&		$ \varepsilon~M_0 $		& 0 \\
\\

$M_{\star} $ & 		$ M_0~\tau'~\varepsilon~(1-R)~ (1- e^{-t/\tau'})  $			
&	0			& $ M_0~\tau'~\varepsilon~(1-R) $ \\
\\

$f_{gas} \equiv \frac{M_{g} }{M_{g} +M_{\star}} $ & 		$ \left[ 1 + \varepsilon~ \tau'~(1-R) ~ (e^{t/\tau'}-1) \right]^{-1} $  	& 	$\infty$  & 0 \\

\\

$Z_{g} \equiv M_{Zg}/M_{g} 	$	&  	$ \frac{p}{\omega} (1- e^{-\omega~\varepsilon~t})$ &
$ 0$ &  $\frac{\varepsilon~p}{\omega}$ \\
\\

\hline

\end{tabular}


\label{table_sol2}
\end{table*}

In summary
\begin{enumerate}
\item{Equilibrium (defined by the requirement $\rm dM_{gas} /dt=0$) bathtub models  predict the SFR to be proportional to the inflow rate.}
\item{Non-equilibrium bathtub models tend to equilibrium at late times. These models can naturally reproduce the fundamental metallicity relation (i.e. imply a SFR dependance in the mass-metallicity relation).}
\item{The cosmological accretion rate, as calculated using semianalytics or cosmological simulations, is slowly varying over most of the age of the Universe. Hence a constant gas inflow rate is generally a good assumption for the purpose of chemical evolution modelling. In a constant inflow model, the relation between metallicity and gas fraction depends only on the outflow loading factor and not on the inflow rate.}  
\item{Non-equilibrium models which fix the inflow rate to be proportional to the SFR implicitly assume the inflow rate to be exponentially declining. This is not a good representation of the cosmological accretion rate for `main sequence' star forming galaxies, but could constitute a good model for strangulation, as in \cite{Peng2015}. Moreover in this model the relation between the metallicity and the gas fraction depends on both the outflow loading factor and the proportionality factor between inflow rate and SFR, causing a degeneracy between inflows and outflows when interpreting metallicity data.}
\end{enumerate}

\end{document}